\documentclass[a4paper,12pt]{article}

\usepackage{amsmath}
\usepackage{amssymb}
\usepackage{enumitem}
\usepackage{amsthm} 
\usepackage{graphicx}
\usepackage{dsfont}
\usepackage[font=small,labelfont=bf]{caption}
\usepackage{fancyhdr}

\def\bSig\mathbf{\Sigma}

\newcommand{\tr}{\mbox{tr}}
\newcommand{\ts}{^\top}  
\newcommand{\diag}{\mathop{\mathrm{diag}}}
\newcommand{\E}{\mathbb{E}}
\newcommand{\Va}{\mathbb{V}}
\newcommand{\bone}{\mathbf{1}} 
\newcommand{\bbeta}{\boldsymbol{\beta}} 
\newcommand{\bmeta}{\boldsymbol{\eta}} 
\newcommand{\bdelta}{\boldsymbol{\delta}} 
\newcommand{\bDelta}{\boldsymbol{\Delta}} 
\newcommand{\blambda}{\boldsymbol{\lambda}} 
\newcommand{\bmu}{\boldsymbol{\mu}} 
\newcommand{\bnu}{\boldsymbol{\nu}} 
\newcommand{\bsigma}{\boldsymbol{\sigma}} 
\newcommand{\be}{\boldsymbol{e}} 
\newcommand{\bg}{\boldsymbol{g}} 
\newcommand{\bD}{\mathbf{D}} 
\newcommand{\bH}{\mathbf{H}} 
\newcommand{\bM}{\mathbf{M}} 
\newcommand{\bQ}{\mathbf{Q}} 
\newcommand{\bS}{\mathbf{S}} 
\newcommand{\bT}{\boldsymbol{T}} 
\newcommand{\bV}{\mathbf{V}} 
\newcommand{\bx}{\boldsymbol{x}} 
\newcommand{\bX}{\mathbf{X}} 
\newcommand{\by}{\boldsymbol{y}} 
\newcommand{\bY}{\boldsymbol{Y}} 
\newcommand{\td}{\mathrm{d}} 
\newcommand{\mbbR}{\mathbb{R}} 
\usepackage{amsmath}
\DeclareMathOperator*{\argmax}{arg\,max}
\DeclareMathOperator*{\argmin}{arg\,min}

\usepackage[figuresright]{rotating}
\usepackage{pbox} 

\newtheorem{theorem}{Theorem}

\theoremstyle{remark} 
\newtheorem{remark}{Remark} 

\usepackage[sort&compress,comma,authoryear]{natbib}

\title{Robust Fitting for Generalized Additive Models for Location, Scale and Shape}
\author{William H.\ Aeberhard$^{1}$, Eva Cantoni$^{2}$,\\Giampiero Marra$^{3}$, and Rosalba Radice$^{4}$\\
   \small$^{1}$Department of Mathematical Sciences, Stevens Institute of Technology, USA\\
   \small$^{2}$Research Center for Statistics and GSEM, University of Geneva, Switzerland\\
   \small$^{3}$Department of Statistical Science, University College London, UK\\
   \small$^{4}$Cass Business School, City, University of London, UK}
\date{}

\begin{document}




\maketitle


\begin{abstract} 
The validity of estimation and smoothing parameter selection for the wide class of generalized additive models for location, scale and shape (GAMLSS) relies on the correct specification of a likelihood function. Deviations from such assumption are known to mislead any likelihood-based inference and can hinder penalization schemes meant to ensure some degree of smoothness for non-linear effects. We propose a general approach to achieve robustness in fitting GAMLSSs by limiting the contribution of observations with low log-likelihood values. Robust selection of the smoothing parameters can be carried out either by minimizing information criteria that naturally arise from the robustified likelihood or via an extended Fellner-Schall method. The latter allows for automatic smoothing parameter selection and is particularly advantageous in applications with multiple smoothing parameters. We also address the challenge of tuning robust estimators for models with non-linear effects by proposing a novel median downweighting proportion criterion. This enables a fair comparison with existing robust estimators for the special case of generalized additive models, where our estimator competes favorably. The overall good performance of our proposal is illustrated by further simulations in the GAMLSS setting and by an application to functional magnetic resonance brain imaging using bivariate smoothing splines.
\end{abstract}


\section{Introduction}
\label{s:intro}  

Generalized additive models for location, scale and shape (GAMLSS) are flexible non-parametric regression models that have been introduced by \cite{RigbyStasinopoulos2005}; see also the recent book and tutorial by \citet{stasinopoulos2017flexible} and \cite{stasinopoulos2018gamlss} for a review. These models allow to use explanatory variables not only to model the location parameter (e.g., the mean) of a response distribution, like in generalized additive models \citep[GAM;][]{hastie1990}, but also the scale and shape parameters. GAMLSSs also go beyond the exponential family of distributions. In fact, the approach can be seen more broadly as a way to model any parameter of any given distribution. As such, some authors refer to it as distributional or multi-parameter regression \citep[e.g.,][]{burke2017multi,lang2014multilevel,pan2003modelling,stasinopoulos2018gamlss}. Software availability for a wide range of families of distributions, such as the \texttt{R} package \texttt{gamlss} \citep{Rpackage_gamlss}, have helped making these models very popular and widely applied in several fields: we can cite \cite{glasbey2009efficiency} (normalizing cDNA microarray), \cite{rudge2005excess} (health impact of temperatures in dwellings), \cite{de2010hands} (long-term survival models for clinical studies), \cite{beyerlein2008alternative} (childhood obesity), and \cite{cole2009age} (charts for child growth curves).
 
The motivation of this paper comes from challenging applications like the one presented in Section~\ref{s:data}. The study first reported in \cite{Landau2003} investigates differences in the brain physiological response to controlled stimuli between anatomically distinct regions. The brain activity response is measured at voxels in a brain slice (a 2D raster image) with sole explanatory variables being the coordinates identifying the location of each voxel. The measurements are highly noisy but the mean response level and its spread are believed to vary smoothly over the brain slice, thus prompting non-linear effects for both location and scale parameters. \citet[p.~329]{Wood17} identified two voxel responses in these data that were deemed too extreme, and were then discarded for the subsequent analysis. We believe a robust fitting of a GAMLSS is hence appropriate here, and this for two reasons: to guarantee that estimates and uncertainties are reliable, and to identify potentially outlying observations in an automated way thanks to robustness weights.


The fitting of GAMLSSs is typically performed by penalized maximum likelihood (ML) estimation. For datasets like the one above, where extreme observations likely occur, the ML estimation procedure suffers from a lack of robustness, meaning that the estimated smooth functions can be distorted by the outliers. Both the nonparametric function estimates themselves and the choice of the smoothing parameters associated to them are affected. To address these issues, we introduce a general robust estimator for GAMLSSs. Our approach covers special cases where robustness has been previously addressed, in particular in the (extended) GAM context \citep{alimadad2011outlier,wong2014robust,croux2012robust}. These works, however, cannot be extended to the more general setting of GAMLSS. Specifically, in contrast with the cited literature which acts at the level of the score equations, we introduce robustness by modifying the objective function following an idea introduced by \cite{EguchiKano2001}. We also propose a novel and general procedure to tune the robustness parameter associated with the robust approach. This problematic issue has been partially ignored in the literature for robust (extended) GAMs. For the selection of the smoothing parameters, we additionally propose robust versions of the Akaike Information Criterion (AIC) and Bayesian Information Criterion (BIC), that can be typically minimized in a grid search, and an adaptation of the Fellner-Schall automatic multiple smoothing parameter selection method \citep{wood2017fellnerschall}, which has important practical advantages. The proposed robust models can be easily used via the newly-revised \texttt{gamlss()} function in the \texttt{R} package \texttt{GJRM} \citep{Rpackage_GJRM}.


In Section~\ref{s:gamlssframework} we introduce the GAMLSS framework and the related estimation procedure which is based on penalized maximum likelihood. Our proposal is fully introduced in Section~\ref{s:robustGAMLSS}, with subsections devoted to the definition of a penalized robust objective function, theoretical properties and inference, the practical implementation of the estimation procedure, smoothing parameter selection, and the challenge of the choice of the robustness tuning constant. In Section~\ref{s:simul} we present two simulation studies to highlight the good behavior of our proposal: one in the GAMLSS setting with a design mimicking the brain imaging data example, and one in the special case of a GAM to allow comparison with existing robust procedure in this context. The brain imaging data analysis is then presented in Section~\ref{s:data}, while conclusions are given in Section~\ref{s:discu}.


\section{GAMLSS Framework and Penalized Estimation}
\label{s:gamlssframework}

\subsection{Framework and Notation}
\label{ss:notation}

Given a sequence of $n$ independent response random variables $Y_1,\ldots,Y_n$, the generalized additive model for location, scale and shape \citep[GAMLSS;][]{RigbyStasinopoulos2005} is defined by
\begin{equation}\label{eq:gamlss}
\begin{aligned}
  Y_i &\sim D(\mu_i, \sigma_i, \nu_i), \ i=1,\ldots,n  \\
  \eta_{1i} &= g_1\left(\mu_i\right)    = \beta_{10} + s_{11}(\bx_{11i}) + \ldots + s_{1 k}(\bx_{1 k i})+ \ldots + s_{1 K_1}(\bx_{1K_1i}), \\
  \eta_{2i} &= g_2\left(\sigma_i\right) = \beta_{20} + s_{21}(\bx_{21i}) + \ldots + s_{2 k}(\bx_{2 k i})+ \ldots + s_{2 K_2}(\bx_{2K_2i}),  \\
  \eta_{3i} &= g_3\left(\nu_i\right)    = \beta_{30} + s_{31}(\bx_{31i}) + \ldots + s_{3 k}(\bx_{3 k i})+ \ldots + s_{3 K_3}(\bx_{3K_3i}), 
\end{aligned}
\end{equation}
where $D$ denotes a family of distributions canonically parametrized in terms of location $\mu_i$, scale $\sigma_i$ and shape $\nu_i$ which are related to the respective predictors $\eta_{di}$ via specified link functions $g_d$, for $d=1,2,3$, $\beta_{d0}\in\mbbR$ are overall intercepts, $\bx_{d k i}$ denotes the $k$th sub-vector of covariates pertaining to term $d$ and observation $i$ (which includes binary, categorical, and continuous variables), and the $K_d$ functions $s_{d k}(\cdot)$ represent generic effects of covariates (linear or not). The distributional assumption of $Y_i$ is understood to be conditional on all covariates. We approximate each $s_{d k}(\bx_{d k i})$ by a linear combination of $J_{d k}$ basis functions $b_{d k j}(\bx_{d k i})$ and regression coefficients $\beta_{d k j}\in\mbbR$ \citep[e.g.,][]{Wood17}
\begin{equation*} 
s_{d k}(\bx_{d k i}) \approx \sum_{j=1}^{J_{d k}} \beta_{d k j} b_{d k j}(\bx_{d k i}).
\end{equation*}
This allows the model summarized by (\ref{eq:gamlss}) to be written in a compact form for the random vector $\bY = (Y_1,\ldots,Y_n)\ts$ as $\bY \sim D(\bmu, \bsigma, \bnu)$ by some slight abuse of notation, where the parameter vectors $\bmu=(\mu_1,\ldots,\mu_n)\ts$, $\bsigma=(\sigma_1,\ldots,\sigma_n)\ts$ and $\bnu=(\nu_1,\ldots,\nu_n)\ts$ are modeled through 
\begin{equation}\label{eq:gamlss2}
\begin{aligned}
  \bmeta_{1} &= g_1(\bmu)    = \bone_n\beta_{10} + \bX_{11}\bbeta_{11}+\ldots+\bX_{1K_1}\bbeta_{1K_1}  = \bX_1\bbeta_1,\\
  \bmeta_{2} &= g_2(\bsigma) = \bone_n\beta_{20} + \bX_{21}\bbeta_{21}+\ldots+\bX_{2K_2}\bbeta_{2K_2}  = \bX_2\bbeta_2,\\
  \bmeta_{3} &= g_3(\bnu)    = \bone_n\beta_{30} + \bX_{31}\bbeta_{31}+\ldots+\bX_{3K_3}\bbeta_{3K_3}  = \bX_3\bbeta_3,
\end{aligned}
\end{equation}
where the functions $g_d$ are applied element-wise, $\bone_n$ is an $n$-dimensional vector of ones, the $(n \times J_{d k})$ matrix $\bX_{d k}$ has $(i,j)$th element $b_{d k j}(\bx_{d k i})$, and $\bbeta_{d k} = (\beta_{d k 1}, \ldots, \beta_{d k J_{k d}})\ts$. The predictors can thus be re-written as $\bmeta_d=\bX_d \bbeta_d$, where $\bX_d = (\bone_n, \bX_{d 1}, \ldots, \bX_{d K_d})$ and $\bbeta_d = (\beta_{d 0}, \bbeta_{d 1}\ts, \ldots, \bbeta_{d K_d}\ts)\ts$. We note that our results and methods here are understood in a fixed-knot framework, i.e.\ that the number of basis functions is fixed at a high value so that any approximation bias in $s_{d k}(\bx_{d k i})$ is negligible compared to estimation variability \citep[as in e.g.][]{vatter2015}.

To enforce a certain degree of smoothness for every approximated $s_{d k}(\cdot)$ function, each $\bbeta_{d k}$ has an associated quadratic penalty $\lambda_{d k} \bbeta_{d k}\ts \bD_{d k} \bbeta_{d k}$, where $\bD_{d k}$ only depends on the choice of basis functions. The smoothing parameter $\lambda_{d k} \in [0,\infty)$ controls the trade-off between fit and smoothness and plays a crucial role in determining the shape of the estimated $\hat{s}_{d k}(\cdot)$. The overall penalty can be written as $\bbeta_d\ts \bD_d \bbeta_d$, where $\bD_d = \diag(0,\lambda_{d 1}\bD_{d 1}, \ldots, \lambda_{d K_d}\bD_{d K_d})$. Following \cite{Wood17}, the approximated $s_{d k}(\cdot)$ smooth functions are subject to centering constraints to ensure identifiability. Examples of smooth function specification include one-dimensional, multi-dimensional, random field and random effect smoothers; see e.g.\ \cite{Wood17} for details. Note that we have considered distributions with up to three parameters (location, scale and shape), hence the adopted notation with $d=1,2,3$, yet the proposed framework can be conceptually extended to distributions with more parameters in a straightforward manner. The families of distributions implemented in this work are listed in Table~S1 in Web Appendix~D.

\subsection{Penalized Log-likelihood}
\label{ss:penloglik}

Let $\bdelta = (\bbeta_{1}\ts,\bbeta_{2}\ts,\bbeta_{3}\ts)\ts \in \bDelta \subseteq \mbbR^{p}$ denote the full model parameter vector. Given a sample of $n$ realizations $y_1,\ldots,y_n$, the log-likelihood function corresponding to (\ref{eq:gamlss2}) is given by
\begin{equation}\label{eq:loglik}
\ell(\bdelta) = \sum_{i=1}^n \ell(\bdelta)_i  = \sum_{i=1}^n \log f\left(y_{i}|\mu_{i}, \sigma_{i}, \nu_{i}\right),
\end{equation}
where $f\left(y_{i}|\cdot\right)$ can either denote the probability density function (pdf) or the probability mass function (pmf) corresponding to the distribution $D$. Because of the flexibility of the smooth terms, the use of an unpenalized optimization algorithm is likely to result in unduly wiggly estimates \citep[e.g.][]{Wood17}. Estimation is thus typically performed by maximizing the penalized version $\ell_p(\bdelta) = \ell(\bdelta) - \frac{1}{2} \bdelta\ts \bS \bdelta$, where $\bS = \diag(\bD_{1},\bD_{2},\bD_{3})$. The smoothing parameters contained in the $\bD_d$'s make up the vector $\blambda = (\blambda_{1}\ts,\blambda_{2}\ts,\blambda_{3}\ts)\ts$. Estimation of $\bdelta$ is typically achieved for a given value of $\blambda$, while the selection of $\blambda$ is often performed by minimizing some prediction error criterion, either as an outer optimization or in an alternating scheme \citep{Wood17}. Examples of such a criterion include cross-validation \citep[CV; e.g.][]{hastie1990} and generalized cross-validation \citep{craven1979} estimates of prediction error, as well as estimates of the Kullback-Leibler divergence between a true model and the fitted one such as the AIC, and the Generalized Information Criterion (GIC) of \citet{konishi1996}.

\section{Robust Estimation}
\label{s:robustGAMLSS}

The estimation procedures mentioned in the previous section rely on strict distributional assumptions. These methods are known to be highly sensitive to deviations from model assumptions \citep[e.g.][]{hampel1986,huber2009}. To this end, we propose a general robust fitting approach, which is valid for the entire class of GAMLSS and that directly yields robust criteria for the selection of smoothing parameters. 

\subsection{Penalized Robustified Log-likelihood}
\label{ss:robpenloglik}

Based on the $\Psi$-divergence approach of \citet{EguchiKano2001}, we introduce the robustified log-likelihood
\begin{equation*}
\tilde{\ell}(\bdelta) = \sum_{i=1}^n \rho_c\big(\ell(\bdelta)_i\big) - b_{\rho}(\bdelta),
\end{equation*}
where, for a given $\bdelta$, the user-specified $\rho_c$ function is designed to reduce low log-likelihood values $\ell(\bdelta)_i$ while leaving large log-likelihood values essentially unchanged, and
\begin{equation}\label{eq:fcct}
b_{\rho}(\bdelta) = \sum_{i=1}^n b_{\rho}(\bdelta)_i = \sum_{i=1}^n  \int \rho_c^\star\big(\log f(y|\mu_{i},\sigma_{i},\nu_{i})\big) \, \td y
\end{equation}
is a correction term ensuring Fisher consistency (see Theorem~\ref{th:asnorm} below), where $\rho_c^\star$ is directly derived from the specified $\rho_c$ through
\begin{equation*}
\rho_c^\star(z) = \int_{-\infty}^z \exp(s) \rho_c'(s) \, \td s,
\end{equation*}
where $\rho_c'(s) = \partial \rho_c(s) / \partial s $. The $\rho$ function is indexed by a so-called robustness tuning constant $c>0$ which regulates the trade-off between loss of estimation efficiency, should the data exactly come from the assumed GAMLSS, and the magnitude of the maximum estimation bias should the data not come from the postulated model. For any given $c$, $\rho_c$ is assumed to be convex, monotonically increasing and twice continuously differentiable over $\mbbR$ and have bounded first derivative $\rho_c'$ within $[0,1]$. The latter can be interpreted as a multiplicative robustness weight, as one would do when weighting the estimating equations in robust $M$-estimation. The important difference here is that the ``robustification'' happens at the log-likelihood level and not by directly applying weights at the score level, such as in \citet{wong2014robust} for example. An advantage of our approach is that it leads to a natural definition of robust criteria for the selection of smoothing parameters (see Section~\ref{ss:robinference}), for instance.


\citet{EguchiKano2001} proposed the following log-logistic $\rho$ function:
\begin{equation*}
\rho_c(z) = \log\frac{1+\exp(z+c)}{1+\exp(c)}, \quad c>0,
\end{equation*}
with corresponding $\rho_c^\star(z) = \exp(z) - \exp(c)\log\big(1+\exp(z+c)\big)$ and first derivative $\rho_c'(z) = \exp(z+c)/\big(1+\exp(z+c)\big)$. Web Figure~S1 in Web Appendix~C displays the log-logistic $\rho_c$ and its first derivative. It illustrates how a smaller value of $c$ leads to an earlier flattening of the $\rho$ function applied on log-likelihood contributions, thus limiting earlier their impact. Note that $\lim_{c\rightarrow\infty}\rho_c(z) = z$ so that an increasingly large $c$ value leads to the (non-robust) original $\ell(\bdelta)$. We discuss the choice of $c$ in Section~\ref{ss:choicetuning}.

For a given smoothing parameter $\blambda$, we define our robust estimator $\hat{\bdelta} = \hat{\bdelta}(\blambda)$ by maximizing the penalized robustified log-likelihood
\begin{equation}\label{eq:robpenloglik}
\hat{\bdelta} = \argmax_{\bdelta} \tilde{\ell}_{p}(\bdelta) = \argmax_{\bdelta}\left\{\tilde{\ell}(\bdelta) - \frac{1}{2} \bdelta\ts \bS \bdelta\right\},
\end{equation}
where the penalty is identical to that of non-robust penalized estimation. Indeed, our robustification scheme targets only deviations in the response variable, the latter which does not appear in $\bdelta\ts \bS \bdelta$ so that only contributions to the unpenalized log-likelihood $\ell(\bdelta)$ need to be accounted for. The robust estimator is thus the solution in $\bdelta$ to the following estimating equations (first-order conditions):
\begin{align}\label{eq:esteq}
\mathbf{0} &= \frac{\partial \tilde{\ell}(\bdelta)}{\partial \bdelta} - \bS\bdelta = \sum_{i=1}^n \rho_c'\big(\ell(\bdelta)_i\big)\frac{\partial \ell(\bdelta)_i}{\partial \bdelta} - \frac{\partial b_{\rho}(\bdelta)}{\partial \bdelta} - \bS\bdelta.
\end{align}
In (\ref{eq:esteq}), the response variable $Y_i$ only appears through $\ell(\bdelta)_i$ since $b_{\rho}$ is an expectation. Thus $\rho_c'$ indeed plays the role of a multiplicative weight within $[0,1]$ which limits the impact of potentially deviating observations given some $\bdelta$. This robustness weight is proved useful both for selecting $c$ (see Section~\ref{ss:choicetuning}) and as a diagnostic tool (see the data analysis in Section~\ref{s:data}).

\subsection{Asymptotic Properties and Inference}
\label{ss:robinference}

The unpenalized robust estimator which maximizes $\tilde{\ell}(\bdelta)$ admits a statistical $M$-functional representation $\bT(F)$, for some generic probability distribution $F$, which is the solution in $\bdelta$ to $\E\left[\psi(Y,\bdelta)\right] = \mathbf{0}$ where
\begin{equation}\label{eq:Tfunctional}
\psi(Y,\bdelta) = \rho_c'\big(\log f(Y|\mu,\sigma,\nu)\big)\frac{\partial \log f(Y|\mu,\sigma,\nu)}{\partial \bdelta} - \frac{\partial }{\partial \bdelta}\E\left[ \rho_c^\star\big(\log f(Y|\mu,\sigma,\nu)\big) \right]
\end{equation}
with expectations taken under $F$. Thus, the finite-sample solution in $\bdelta$ to
\begin{equation*}
\frac{1}{n}\sum_{i=1}^n \left\{\rho_c'\big(\ell(\bdelta)_i\big)\frac{\partial \ell(\bdelta)_i}{\partial \bdelta} - \frac{\partial b_{\rho}(\bdelta)_i}{\partial \bdelta}\right\} = \mathbf{0}
\end{equation*}
can be written as $\bT(F_n)$, where $F_n$ denotes the empirical distribution putting mass $1/n$ on each observation. $\bT(F_n)$ amounts to an unpenalized robust estimator.

To discuss the asymptotic properties of the proposed (penalized) robust estimator, we define $\bdelta_0$ as the parameter value to which the unpenalized MLE maximizing $\ell(\bdelta)$ in (\ref{eq:loglik}) converges, as $n\rightarrow\infty$. By viewing $\bdelta_0$ as the ``true'' parameter that generates the data under distribution $D$ with moments defined in Equation~(\ref{eq:gamlss2}), Theorem~\ref{th:asnorm} below establishes the Fisher consistency of $\hat{\bdelta}$ and its asymptotic distribution; the proof is deferred to Web Appendix~A.

\begin{theorem}\label{th:asnorm}
Under conditions (\text{C}1)--(\text{C}5) in Web Appendix~A, as $n\rightarrow\infty$ the penalized robust estimator $\hat{\bdelta}$ admits the same $M$-functional representation $\bT$ as the unpenalized robust estimator and we have $\bT(D) = \bdelta_0$. Moreover, $\sqrt{n}(\hat{\bdelta}-\bdelta_0) \underset{n\rightarrow\infty}{\overset{d}{\longrightarrow}} \text{N}(\mathbf{0}, \bV(\bdelta_0))$,
where the asymptotic covariance matrix is given by the so-called sandwich formula $\bV(\bdelta) = \bM(\bdelta)^{-1} \bQ(\bdelta) \bM(\bdelta)^{-\sf T}$, where
\begin{align}\label{eq:robunpenMQ}
\bM(\bdelta) = -\E\left[ \frac{\partial^2 \tilde{\ell}(\bdelta)}{\partial\bdelta\partial\bdelta\ts} \right] \quad \text{and} \quad \bQ(\bdelta) = \E \left[\left(\frac{\partial \tilde{\ell}(\bdelta)}{\partial\bdelta} \right) \left(\frac{\partial \tilde{\ell}(\bdelta)}{\partial\bdelta} \right)\ts \right],
\end{align}
with expectations taken under the assumed distribution $D$.
\end{theorem}

In Theorem~\ref{th:asnorm}, $\bT(D) = \bdelta_0$ means that $\hat{\bdelta}$ is Fisher consistent: it returns the true parameter when $\bT$ is evaluated at the assumed distribution $D$, which implies that $\hat{\bdelta}$ is asymptotically unbiased for $\bdelta_0$. The influence function \citep[IF;][]{hampel1974influence} of the Fisher consistent functional $\bT$ is proportional to the score $\psi(Y,\bdelta)$ given in (\ref{eq:Tfunctional}). This score being bounded in the response variable $Y$ thanks to $\rho_c' \in [0,1]$, the IF is itself bounded. This guarantees a bounded maximum asymptotic bias under arbitrary contamination in $Y$, which is the main robustness property of $\hat{\bdelta}$. 

\begin{remark}
The asymptotic variance $\bV(\bdelta_0)$ in Theorem~\ref{th:asnorm} corresponds to an unpenalized robust estimation because we assume the usual asymptotically vanishing penalty for consistency (see condition (C5) in Web Appendix~A). A better approximation of the finite-sample covariance matrix with non-zero penalty can be obtained from a Taylor expansion of the penalized robustified score, as given in Equation~(2) in Web Appendix~A. It amounts to $\bV_p(\bdelta_0) = \bM_p(\bdelta_0)^{-1} \bQ(\bdelta_0) \bM_p(\bdelta_0)^{-\sf T}$, where $\bM_p(\bdelta) = -\E\left[ \frac{\partial^2 \tilde{\ell}_{p}(\bdelta)}{\partial\bdelta\partial\bdelta\ts} \right] = \bM(\bdelta) + \bS$. In these expressions, $\bdelta_0$ being unknown in practice one would typically ``plug-in'' the estimate $\hat{\bdelta}$ to compute standard errors. This allows for the computation of approximate (point-wise) confidence intervals, which can then be interpolated for confidence bands for non-linear effects. See for instance \citet[][p.~39]{croux2012robust} for the analogue in the extended GAM setting.
\end{remark}


\begin{remark}\label{rem:Bayescov}
An alternative covariance can be computed following an empirical Bayes approach, which is often reported to lead to good finite-sample coverage of confidence intervals in the frequentist sense \citep[see e.g.][]{MW12sc,Wood17}. For a given $\blambda$, viewing the quadratic penalty as an improper Gaussian prior distribution for $\bdelta$ (seen as a random vector here), with mean zero and covariance $\bS^{-1}$, the joint density of $(\bY,\bdelta)$ is given, up to normalization constants, by $L(\by,\bdelta;\blambda) = \exp\big(\tilde{\ell}(\bdelta)\big) \exp\big(-\bdelta^\top \bS \bdelta/2\big)|\bS|^{1/2}$, with $|\cdot|$ denoting matrix determinant. We seek the covariance of the posterior distribution of $\bdelta | \bY$, as the posterior mode corresponds to the robust estimate $\hat{\bdelta}$. As in \citet{wood2017fellnerschall}, a second-order Taylor expansion of the posterior log-density about its mode reveals that as $n\rightarrow\infty$ the posterior distribution approaches a multivariate Gaussian with covariance given by $\bM_p(\hat{\bdelta})^{-1}$. Our experience is that the observed version of this posterior covariance matrix, $\widehat{\bM}_p(\hat{\bdelta})^{-1} = \left(\widehat{\bM}(\hat{\bdelta}) + \bS\right)^{-1}$, where $\widehat{\bM}(\bdelta) = -\frac{\partial^2 \tilde{\ell}(\bdelta)}{\partial\bdelta\partial\bdelta\ts}$, can be used as a computationally efficient alternative to $\bV_p(\hat{\bdelta})$.
\end{remark}

The effective degrees of freedom (edf) of smooth terms are a valuable tool for assessing the degree of smoothness achieved by a fit. We follow the discussion of \citet[][Chapter~6]{Wood17} based on links with generalized linear mixed models and restricted ML estimation to obtain that the edf of a GAMLSS robust fit is $\text{tr}\big\{\widehat{\bM}_p(\hat{\bdelta})^{-1}\widehat{\bQ}(\hat{\bdelta})\big\} = \text{tr}\big\{(\widehat{\bM}(\hat{\bdelta}) + \bS)^{-1}\widehat{\bQ}(\hat{\bdelta})\big\}$, where $\widehat{\bQ}(\bdelta) = \left(\frac{\partial \tilde{\ell}(\bdelta)}{\partial\bdelta}\right) \left(\frac{\partial \tilde{\ell}(\bdelta)}{\partial\bdelta}\right)\ts$. This term matches the ``penalty term'' of our robust AIC introduced in Section~\ref{ss:robsmoothparamsel} below.

\subsection{Estimation Approach and Implementation}
\label{ss:robestimation}

To maximize (\ref{eq:robpenloglik}), we have modified the efficient and stable trust region algorithm of \citet{MA16} to accommodate the robustified objective function and corresponding correction term $b_{\rho}(\bdelta)$. Estimation of $\bdelta$ and $\blambda$ is carried out as follows. At iteration $a$, holding $\blambda$ fixed and for some tuning constant value $c$, for a given $\bdelta^{[a]}$ we maximize equation (\ref{eq:robpenloglik}) using a trust region algorithm \citep{conn}:
\begin{equation}\label{eq1st1}
\bdelta^{[a+1]} = \bdelta^{[a]} + \argmin_{\be:\|\be\| \leq \Delta^{[a]}} \breve{\tilde{\ell}}_p(\be;\bdelta^{[a]}),
\end{equation}
where $\|\cdot\|$ denotes the Euclidean norm, $\Delta^{[a]}$ is the radius of the trust region which is adjusted throughout the iterations, $\breve{\tilde{\ell}}_p(\be;\bdelta^{[a]}) = -\left(\tilde{\ell}_p(\bdelta^{[a]}) + \be\ts\bg_p(\bdelta^{[a]}) + \frac{1}{2}\be\ts\bH_p(\bdelta^{[a]})\be\right)$, $\bg_p(\bdelta^{[a]}) = \bg(\bdelta^{[a]}) - \bS\bdelta^{[a]}$ and $\bH_p(\bdelta^{[a]}) = \bH(\bdelta^{[a]}) - \bS$, and where the vector $\bg(\bdelta^{[a]})$ consists of the stacked $\bg_d(\bdelta^{[a]}) = \partial \tilde{\ell}(\bdelta) / \partial \bbeta_d |_{\bbeta_d = \bbeta_d^{[a]}}$ for $d=1,2,3$, and the Hessian matrix $\bH$ has elements $\bH(\bdelta^{[a]})_{d,h} = \partial^2 \tilde{\ell}(\bdelta)/\partial \bbeta_d \partial \bbeta_h\ts|_{\bbeta_d={\bbeta}_d^{[a]},\bbeta_h={\bbeta}_h^{[a]}}$, for $d,h=1,2,3$.
Equation~(\ref{eq1st1}) uses a quadratic approximation of $-\tilde{\ell}_p$ about $\bdelta^{[a]}$ (the so-called model function) in order to choose the best $\be^{[a+1]}$ within the ball centered in $\bdelta^{[a]}$ of radius $\Delta^{[a]}$, the trust region. Close to the converged solution, the trust region usually behaves like an unconstrained optimization algorithm.

Trust region algorithms have several advantages over classical alternatives. For instance, in line search methods, when an iteration falls in a long plateau region, the search for step $\bdelta^{[a+1]}$ can occur so far away from $\bdelta^{[a]}$ that the evaluation of the model log-likelihood may be indefinite or not finite, in which case the user's intervention is required. Trust region methods, on the other hand, always solve the sub-problem (\ref{eq1st1}) before evaluating the objective function. So, if $\tilde{\ell}_p$ is not finite at the proposed $\bdelta^{[a+1]}$ then step $\be^{[a+1]}$ is rejected, the trust region shrunken, and the optimization computed again. The radius is also reduced if there is no agreement between the model and objective functions (i.e., the proposed point in the region is not better than the current one). Reversibly, if an agreement occurs, the trust region is expanded for the next iteration. In summary, $\bdelta^{[a+1]}$ is accepted if it improves over $\bdelta^{[a]}$ and allows for the evaluation of $\breve{\tilde{\ell}}_p$, whereas the reduction/expansion of $\Delta^{[a+1]}$ is based on the similarity between model and objective functions. Theoretical and practical details of the method can be found in \citet[][Chapter 4]{Nocedal} and \citet{Geyer}. The latter also discusses the necessary modifications to the sub-problem (\ref{eq1st1}) and the radius for ill-scaled variables. 

The analytical score and Hessian of (the non-robust) $\ell(\bdelta)$ can be derived in a modular way. This allows for a direct extension to other families of distributions not included in Table~1 in Web Appendix~D as long as their pdf/pmf are known and their derivatives with respect to their parameters exist. Regarding the optimization of the robustified $\tilde{\ell}_p(\bdelta)$, the integral defining $b_{\rho}(\bdelta)$ in (\ref{eq:fcct}), as well as its derivatives, in general have to be approximated. For discrete distributions over countably infinite supports, this amounts to a straightforward truncation of a converging infinite sum. For continuous distributions, we rely on a unidimensional adaptive Gaussian quadrature rule for which we compute data-based finite bounds for numerical stability and to increased speed.

\subsection{Robust Selection of Smoothing Parameters}
\label{ss:robsmoothparamsel}

Our robustification scheme with $\rho_c$ directly applied on log-likelihood contributions has the advantage of yielding a natural robust AIC (RAIC). Following the construction of the generalized information criterion (GIC) of \citet{konishi1996}, we can define the Kullback-Leibler divergence $d_{\text{KL}}$ between the true distribution $G$ that generated the data, with density $g$, and the distribution corresponding to our robustified likelihood (up to normalization constants) as
\begin{equation}\label{eq:dkl}
d_{\text{KL}} = \E_G \big[ \log \big(g(Y)/\exp(\tilde{\ell}(\bdelta,Y))\big) \big] = \E_G[\log g(Y)] - \E_G[\tilde{\ell}(\bdelta,Y)],
\end{equation}
where $\tilde{\ell}(\bdelta,Y) = \rho_c\big(\log f(Y|\mu,\sigma,\nu)\big) - \int \rho_c^\star\big(\log f(y|\mu,\sigma,\nu)\big) \, \td y$. The  generic random variable $Y$ here stands for an out-of-sample observation to be predicted, thus $d_{\text{KL}}$ represents a measure of prediction error. Minimizing $d_{\text{KL}}$ with respect to $\bdelta$ is equivalent to maximizing $\E_G[\tilde{\ell}(\bdelta,Y)]$ since the first term on the right hand side of (\ref{eq:dkl}) is a constant. But because $G$ is unknown, the estimator $(1/n)\sum_{i=1}^n \tilde{\ell}(\bdelta,Y_i)$ is used, which is biased for $\E_G[\tilde{\ell}(\bdelta,Y)]$. In the GIC framework, the first-order correction of this bias depends on the estimator used for $\bdelta$. We consider here the penalized robust estimator $\hat{\bdelta}$, so that by Theorem~2.2 of \citet{konishi1996} the bias correction amounts to $\text{tr}\big\{\bM_p(\bdelta)^{-1}\bQ(\bdelta)\big\} = \text{tr}\big\{(\bM(\bdelta) + \bS)^{-1}\bQ(\bdelta)\big\}$. Thus we define the RAIC as
\begin{equation}\label{eq:raic}
\text{RAIC}(\blambda) = -2 \tilde{\ell}(\bdelta) + 2\text{tr}\big[(\widehat{\bM}(\bdelta) + \bS)^{-1}\widehat{\bQ}(\bdelta)\big],
\end{equation}
where recall that $\bS=\bS(\blambda)$, and the observed matrices $\widehat{\bM}(\bdelta)$ and $\widehat{\bQ}(\bdelta)$ allow for fast computations. Selecting $\blambda$ can thus be done by minimizing $\text{RAIC}(\blambda)$. In (\ref{eq:raic}), since all terms are based on the robustified $\tilde{\ell}(\bdelta)$, the RAIC naturally inherits robustness and the selected $\blambda$ is thus expected to remain stable in the presence of model deviations.

Minimizing an AIC-type criterion for smoothing parameter selection is known to favor more complex models, with function estimates more on the wiggly side. As this feature may carry over to our RAIC, an alternative is to consider a robust version of the Bayesian Information Criterion where its heavier penalty coefficient ($\log(n)$ rather than $2$) generally favors simpler models, with smoother function estimates. Similarly to \citet{wong2014robust}, in our setting a robust BIC (RBIC) is naturally given by
\begin{equation*}
\text{RBIC}(\blambda) = -2 \tilde{\ell}(\bdelta) + \log(n)\text{tr}\big[(\widehat{\bM}(\bdelta) + \bS)^{-1}\widehat{\bQ}(\bdelta)\big].
\end{equation*}

That being said, the proposed RAIC and RBIC procedures involve two nested optimizations: an inner optimization for computing $\hat{\bdelta}$ given $\blambda$, and an outer optimization over $\blambda$. The high computational cost involved makes the selection of $\blambda$ nearly unfeasible, or unbearably slow, whenever more than one or two smoothers are considered. We therefore propose an alternative robust selection method that can be automated as part of the estimation process with little computational overhead. This alternative is a robust version of the Fellner-Schall method recently introduced in \citet{wood2017fellnerschall}, which we will call the extended Fellner-Schall (EFS) method. Web Appendix~B provides the detailed development, the main ideas can be summarized as follows. First, we take the empirical Bayes viewpoint as in Remark~\ref{rem:Bayescov} above to consider the quadratic penalty as an improper Gaussian prior on $\bdelta$, resulting in the the joint (robustified) likelihood $L(\by,\bdelta;\blambda)$. Next, we approximate the integral defining the marginal likelihood $L(\by;\blambda) = \int_{\bDelta} L(\by,\bdelta;\blambda) \,\td \bdelta$ by Laplace's method. By considering the estimate $\hat{\bdelta} = \hat{\bdelta}(\blambda)$ as based on a previous iterate for $\blambda$, we obtain a tractable expression for (the Laplace-approximated) $ \partial \log L(\by;\blambda) / \partial \blambda$. Finally, we follow the heuristic reasoning of \cite{wood2017fellnerschall} to derive the following update from iteration $[k]$ to $[k+1]$ for all elements of $\blambda$:
\begin{align*}
\lambda^{[k+1]}_j &= \lambda^{[k]}_j \times \frac{\tr\big\{\bS(\blambda^{[k]})^{-1} \left.\partial \bS(\blambda)/\partial \lambda_j\right|_{\blambda=\blambda^{[k]}}\big\} - \tr\big\{\widehat{\bM}_p(\hat{\bdelta})^{-1} \left.\partial \bS(\blambda)/\partial \lambda_j\right|_{\blambda=\blambda^{[k]}} \big\}}{\hat{\bdelta}^\top \big(\partial \bS(\blambda)/\partial \lambda_j|_{\blambda=\blambda^{[k]}}\big) \hat{\bdelta}},
\end{align*}
where $\hat{\bdelta} = \hat{\bdelta}(\blambda^{[k]})$ here. In this expression, $\partial \bS(\blambda)/\partial \lambda_j$ is straightforward to write down and implement since $\bS(\blambda)$ is block-diagonal and each block is typically linear in the components of $\blambda$ and only involves the (known) basis functions. We note that under the conditions of Theorem~1 the update guarantees by construction that $\blambda$ remains positive and that the iterates converge whenever the gradient with respect to $\blambda$ gets arbitrarily close to zero. This update rule can thus be alternated with computing $\hat{\bdelta}$ in an automated and efficient way since both rely on similar quantities (see Section~\ref{ss:robestimation}).

\begin{remark}
The proposed EFS method is simple to implement and avoids unfeasible grid searches. All that is required is a set of explicit formulas, as given above, to update $\blambda$ in order to increase the (Laplace-approximated) marginal robustified log-likelihood. Our derivation also highlights the method's broader appeal since it can be easily adapted to modeling situations requiring the use of non-standard models and estimators (i.e.\ beyond the robust estimation in this paper) as long as a Laplace-approximated marginal likelihood is available.
\end{remark}

\subsection{Choice of the Robustness Tuning Constant}
\label{ss:choicetuning}

The robustness tuning constant $c$ regulates how early $\rho_c$ starts to diminish the contribution of an observation to the objective function $\tilde{\ell}$. The choice of $c$ is typically made before fitting the model to data by targeting a certain loss of estimation efficiency with respect to the MLE at the assumed model. With strictly parametric models, the usual criterion is the ratio of the traces of the asymptotic covariance matrices of the model parameters. But with non-parametric models, where basis function coefficients are subject to some smoothness constraint (as is the case here) the asymptotic covariance matrices of the penalized MLE and of the robust estimator are not necessarily comparable. The reason is that robust estimation may achieve a different degree of smoothness, i.e.\ a different bias-variance trade-off stemming from different $\blambda$ values selected by minimizing some prediction error criterion. If the two estimators achieve different degrees of smoothness, then the coefficients variances are not necessarily on the same scale and are thus not comparable. One may constrain the smoothness to be similar between the two estimation methods, but this would defeat the purpose of robustness: we are indeed interested in potential differences between the fitted functions and typically suspect that deviating observations may push classical estimates to be too wiggly. Hence the need for a different criterion for the choice of $c$. We note that previous works \citep{alimadad2011outlier,croux2012robust,wong2014robust} have not discussed this important issue, resorting to somewhat default values for $c$ taken from strictly parametric cases.

We propose a novel general criterion for the selection of the tuning parameter $c$ which covers both additive models and strictly parametric ones. It is simulation-based and relies on the heuristic idea of controlling how the robustness weights at the score level (represented here by $\rho_c'$) behave under data generated from the assumed model. Our procedure is as follows:
\begin{enumerate}
  \item[Step 1:]{For a given tuning constant value $c$, compute the robust estimator $\hat{\bdelta}_c$ on the original data by maximizing (\ref{eq:robpenloglik}), including the optimal smoothing parameter $\hat{\blambda}_c$.
  }
  \item[Step 2:]{For a large number of Monte Carlo replications $B$, for $b \in \{1,\ldots,B\}$ repeat:\begin{enumerate}
      \item{Generate a response vector $\by_b$ given the original design and covariates according to the assumed model in (\ref{eq:gamlss2}) using $\hat{\bdelta}_c$ as generating parameter.
      }
      \item{Use both $\hat{\bdelta}_c$ and $\hat{\blambda}_c$ to compute the vector of robustness weights $(w_{b,1},\ldots,w_{b,n})\ts$, where $w_{b,i} = \rho_c'(\ell(\hat{\bdelta}_c)_{b,i})$ with $\ell(\hat{\bdelta}_c)_{b,i}$ denoting the log-likelihood value corresponding to the $i$th entry in $\by_b$. Compute the sum of the robustness weights $w_b = \sum_{i=1}^n w_{i,b}$.
      }
  \end{enumerate}
  }
  \item[Step 3:]{The criterion corresponding to $c$ is the median downweighting proportion (MDP) over the $B$ independent replicates: $\text{median}\{w_1/n,\ldots,w_B/n\}$.
  }
  \item[Step 4:]{Repeat Steps~1--3 to find the $c$ value matching a target MDP (e.g.\ $\text{MDP}=0.95$).
  }
\end{enumerate}
Since $\rho_c'(\ell(\bdelta)_{b,i}) \in [0,1]$ for any $\bdelta$ by construction, the ratio $w_b/n$ indeed represents how much downweighting has occurred on a particular sample $\by_b$. The value $w_b/n=1$ indicates no downweighting at all, i.e.\ the corresponding estimate is the penalized MLE.

We empirically confirmed over a variety of models (through simulations not presented here) that the MDP indeed increases monotonically with $c$ until reaching one and remaining constant beyond that. This implies that our new criterion shares a one-to-one relation with the traditional criterion of the ratio of the trace of the asymptotic covariance matrices within the subset of $c$ values that lead to some downweighting under the given design. The MDP is not asymptotic and is in effect tailored to the model and design of the data under study. We finally note that no heavy computation is involved in Step~2: we do not estimate parameters on the simulated $\by_b$ vectors, we only need to evaluate the log-likelihood at the true parameter $\hat{\bdelta}_c$ that generated the sample. In addition, our experience is that Monte Carlo simulation variability is quite small in the MDP so that $B=100$ seems sufficient for most practical purposes.

\section{Simulation Studies}
\label{s:simul}

To investigate the finite sample properties of the proposed estimator, we carry out two simulation studies. In the first one, we assess the robustness properties of our methodology in a GAMLSS setting inspired by the motivating data we analyze in Section~\ref{s:data}. In the second simulation study, we compare our proposal to existing alternatives in the simpler setting of a GAM. All computations are performed in \texttt{R} \citep{CRAN}. Our robust estimator is available in the \texttt{R} package \texttt{GJRM} \citep{Rpackage_GJRM}.


\subsection{Simulation under a GAMLSS}
\label{ss:simgamlss}

The motivating brain imaging data introduced in Section~\ref{s:data} have a nonnegative response variable representing the physiological activation level of voxels in a ``brain slice''. There are two covariates, \texttt{X} and \texttt{Y}, defining the location of each voxel. The combinations of \texttt{X} and \texttt{Y} result in a sample size of $n=1567$. In this simulation study, we use the covariates to generate a response for each voxel according to a GAMLSS with a gamma distribution with expectation $\mu$ and variance $\sigma^2\mu^2$ where $\log(\mu) = \eta_1 = s_1(\texttt{X},\texttt{Y})$ and $\log(\sigma) = \eta_2 = s_2(\texttt{X},\texttt{Y})$. The smooth functions $s_1$ and $s_2$ are constructed to mimic the main features of the fitted surfaces on the real data in Section~\ref{s:data}, see Figure~S2 in Web Appendix~C.

To generate data that is contaminated in a similar way to what is observed in the real data, we modify a clean simulated dataset by choosing at random 78 ($=5\%$) of the responses falling in the upper-right corner of the brain slice, for $\texttt{X} > 70$ and $\texttt{Y} > 30$, and by adding 10 to their original value. We simulate 200 replications of the above in both a ``clean'' scenario (at the assumed model) and in the contaminated scenario. For each replication we fit a gamma GAMLSS with log links for both $\mu$ and $\sigma$, both with a classical (ML) and with our robust estimation method. We use bivariate thin plate regression splines with \texttt{k=100} bases to approximate the $s_1$ and $s_2$ smooth functions. Both estimators rely on the EFS method for selecting the smoothing parameters. The robust estimator is tuned to achieve an MDP of 0.95, resulting in $c=3.1$ given the design. We assess estimation performance by investigating the differences between the true parameter and the estimated one, both on the linear predictor scale ($\eta_1$ and $\eta_2$) and on the canonical parameter scale ($\mu$ and $\sigma$). We also compute the mean squared error (MSE) of each target $\theta$ computed as $\text{MSE}(\hat{\theta},\theta) = \frac{1}{n} \sum_{i=1}^n (\hat{\theta}_i - \theta_i)^2$, where $\theta$ is one of $\eta_1$, $\eta_2$, $\mu$ or $\sigma$.

Figure~\ref{fig:simgamlss.mse.eta} below presents boxplots of the MSE of both methods under both scenarios, while Figure~S3 in Web Appendix~C shows the same but with the vertical scales manually set to improve visualization. Similarly, Figures~S4 and S5 present boxplots of MSEs on the scale of $\mu$ and $\sigma$. In the clean data scenario, the MSE of classical estimates for both parameters is slightly smaller than that of robust estimates, as theoretically expected. When the data are contaminated, the MSE of classical estimates explodes whereas the MSE of the robust method only shows a slight increase with somewhat more variability across replications.

\begin{figure}
\centering
\includegraphics[width=\textwidth]{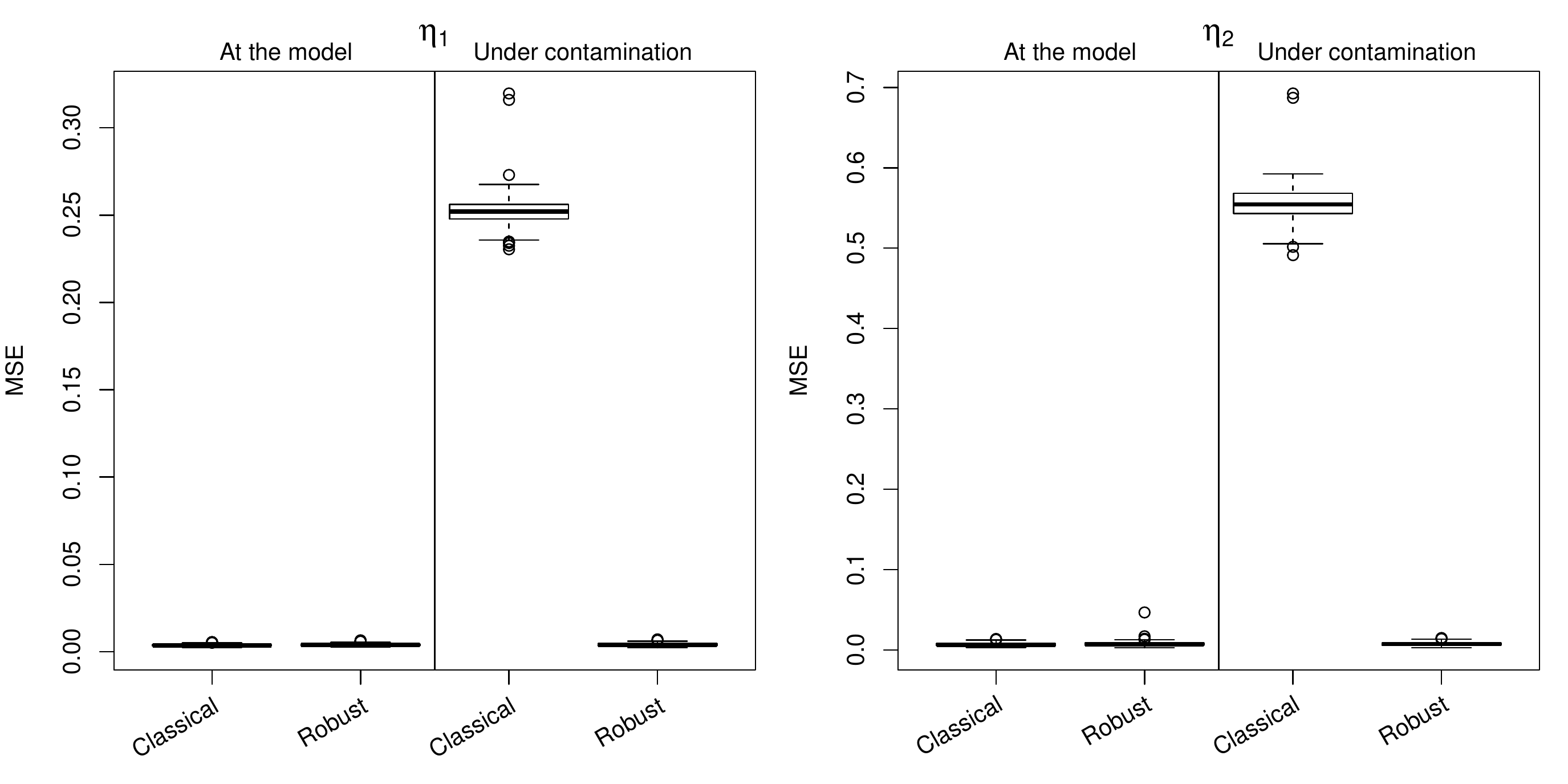}
\caption{GAMLSS simulation, MSE of the linear predictors $\eta_1$ (left panel) and $\eta_2$ (right panel) for classical and robust methods with data generated at the assumed model and under contamination.}
\label{fig:simgamlss.mse.eta}
\end{figure}

\begin{figure}
\centering
\includegraphics[width=\textwidth]{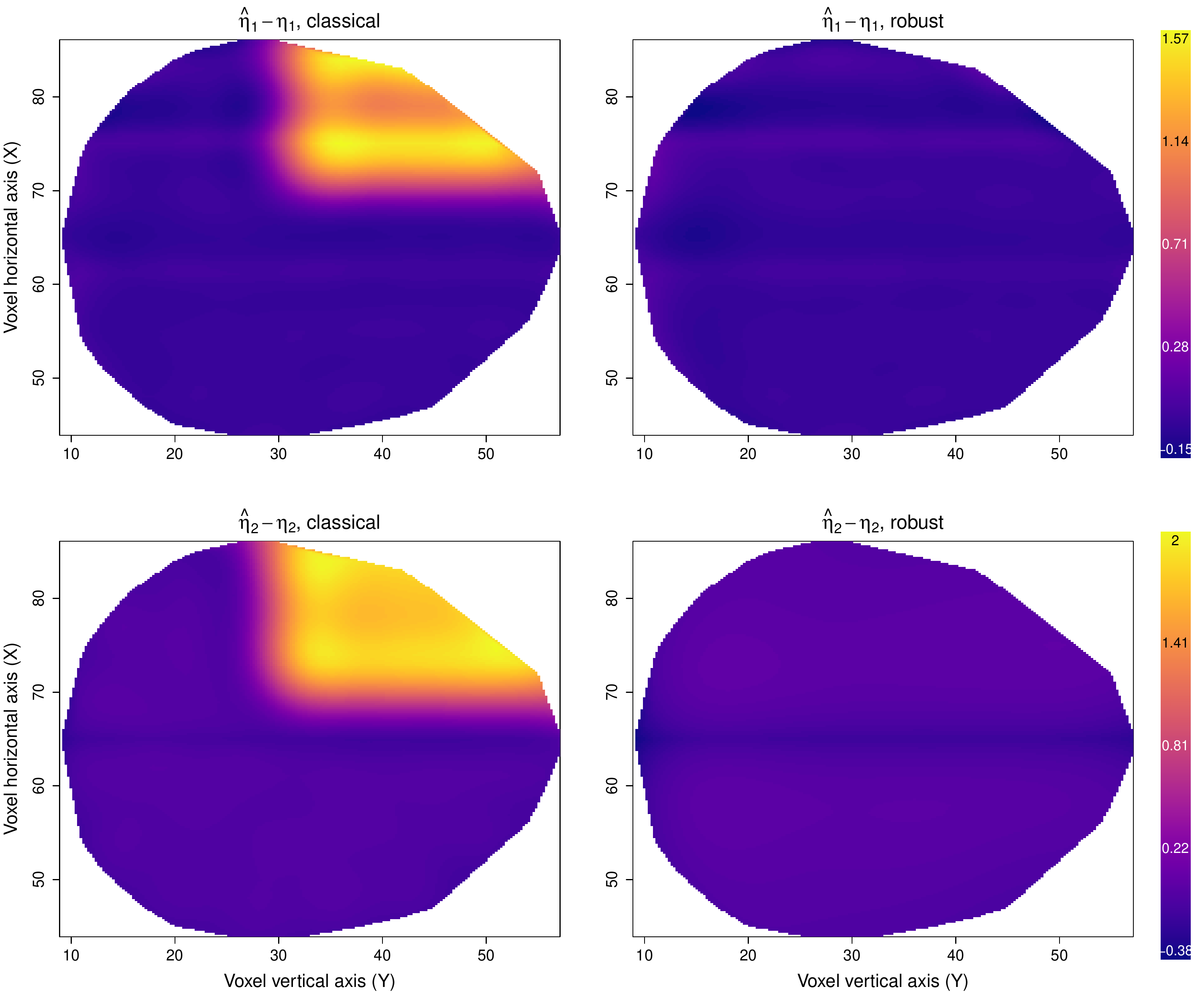}
\caption{GAMLSS simulation, surfaces of the average bias for the linear predictors $\eta_1$ (top row) and $\eta_2$ (bottom row) based on classical (left column) and robust (right column) estimation methods, under contamination.}
\label{fig:simgamlss.biascont}
\end{figure}

We investigate these differences further by looking at the fitted surfaces for $s_1(\texttt{X},\texttt{Y})$ and $s_2(\texttt{X},\texttt{Y})$. Figure~\ref{fig:simgamlss.biasclean} below shows colored surfaces representing the average bias across replications $\frac{1}{200}\sum_{j=1}^{200} (\hat{\theta}_{i,j}-\theta_i)$, where $i=1,\ldots,n$ and $\theta$ is either $\eta_1$ or $\eta_2$, in the clean data scenario; Figure~\ref{fig:simgamlss.biascont} shows the same under the contaminated scenario. Note that the coloring scales are not the same between the two figures. At the assumed model, we see that both methods perform equally well, showing overall little bias centered about zero. However, under contamination the classical estimates show a large positive bias in the top-right corner of the brain slice, which is precisely the area that is contaminated ($\texttt{X} > 70$ and $\texttt{Y} > 30$). Under contamination, the robust estimates show roughly similar biases than at the model, meaning that the fitted surfaces are quite stable in spite of the contamination. In Web Appendix~C, Figures~S6 and S7 present similar colored surfaces but for $\mu$ and $\sigma$; the results are essentially the same. Overall, this simulation study not only highlights the robustness property of our proposed estimator but also how tuning for an MDP of 0.95 yields smooth functions estimations that are nearly indistinguishable from ML-based ones when the data come form the assumed model.

\begin{figure}
\centering
\includegraphics[width=\textwidth]{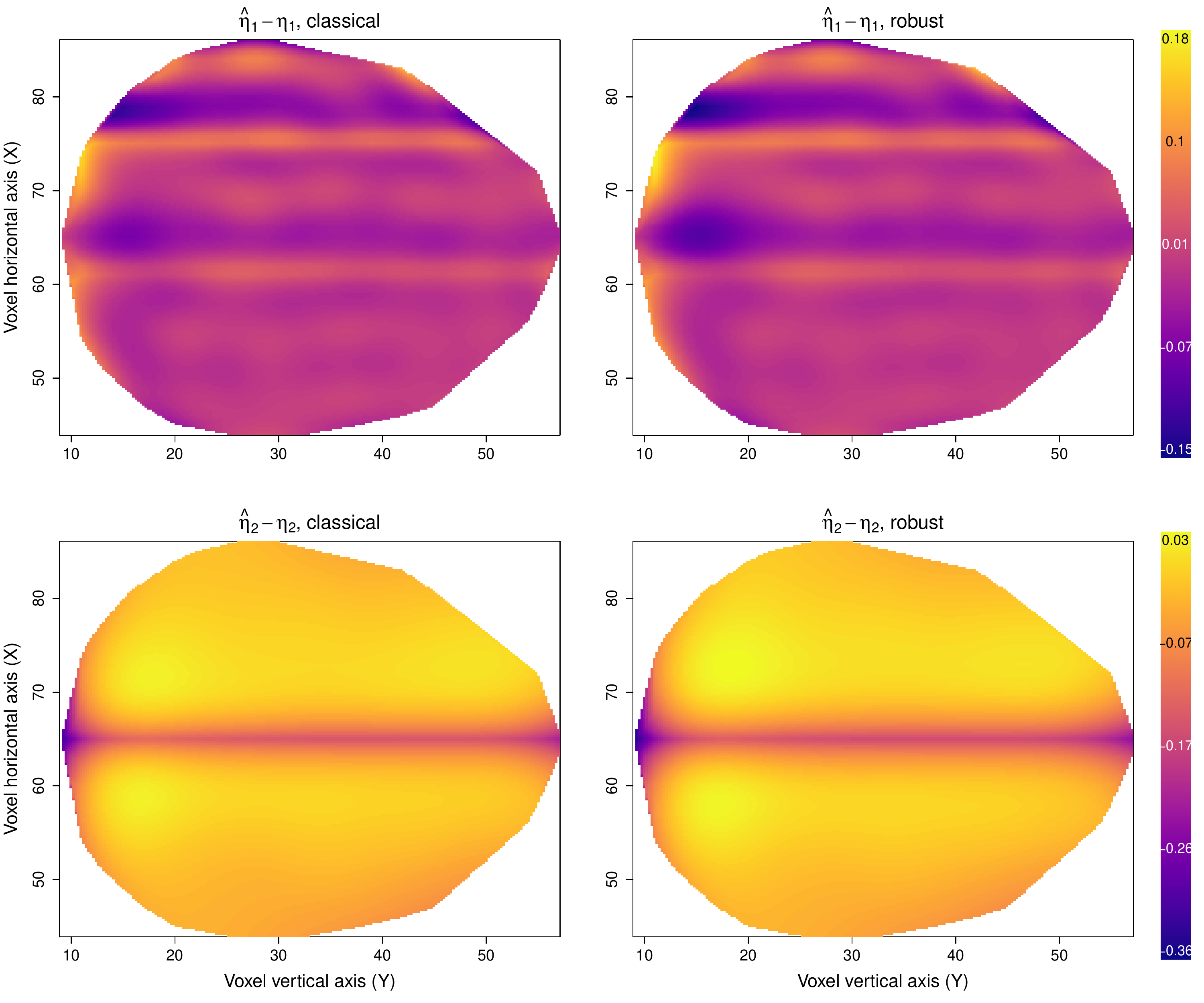}
\caption{GAMLSS simulation, surfaces of the average bias for the linear predictors $\eta_1$ (top row) and $\eta_2$ (bottom row) based on classical (left column) and robust (right column) estimation methods, at the assumed model.}
\label{fig:simgamlss.biasclean}
\end{figure}

\subsection{Comparison to Robust Alternatives in a GAM Setting}
\label{ss:simgam}

In order to compare our proposed estimation method to existing robust approaches in the special case of a GAM, we consider here one of the simulation designs of \citet{wong2014robust}. For $i=1,\ldots,n$, we generate independent responses $Y_i \sim \text{Poisson}(\mu_i)$ with $\mu_i = \exp(\eta_i)$ and $\eta_i = 4 \cos(2\pi(1 - x_i^2))$, where the $x_i$'s are independently drawn from a $\text{Uniform}(0,1)$ distribution. The sample size is set to $n=100$. Following \citet[][p.~280]{wong2014robust}, contaminated data are obtained by randomly selecting $5\%$ of the original responses and changing them to the nearest integer $y_i u_{1}^{u_2}$, where $u_1$ is drawn from a $\text{Uniform}(2,5)$ distribution and where $u_2$ is randomly set to either $1$ or $-1$. We simulate $200$ replications.

We compare the following methods, with the same setting choices as in \citet{wong2014robust}:
\begin{itemize}
  \item{
  AS: the approach of \citet{alimadad2011outlier} with \texttt{span}=0.5;
  }
  \item{
  CGP: the approach of \citet{croux2012robust} with \texttt{nknots}=15;
  }
  \item{
  WYL: the approach of \citet{wong2014robust} with $k=30$ basis functions and with smoothing parameter chosen by minimizing their robust BIC, following their recommendation;
  }
  \item{
  GAMLSS: our proposed approach with $k=20$ basis functions;
  }
  \item{
  Classical: ML-based estimation with $k=20$ basis functions and smoothing parameter selected by the Fellner-Schall method of \citet{wood2017fellnerschall}.
  }
\end{itemize}
All existing approaches build on \citet{cantoni2001} to define robust penalized estimating equations for $\bdelta$. \cite{croux2012robust} additionally define a similar set of estimating equations for the dispersion parameter in their extended GAM setting. That is, all these approaches robustify estimating (score) equations, typically by appending weights, whereas our proposed approach directly robustifies a likelihood. Regarding smoothers and basis functions, \citet{alimadad2011outlier} use local linear fits as smoothers; \citet{croux2012robust} use P-splines; while in \citet{wong2014robust} the nonparametric fits are based on thin plate regression splines. Regarding the smoothing parameter selection, \citet{alimadad2011outlier} use a robust version of CV defined as a sum of squared weighted residuals in line with \citet{cantoni2001resistant}, and implemented it in a ``brute-force'' way; \citet{croux2012robust} construct a robust GCV criterion and a robust AIC by applying some bounded function to the deviances appearing in the classical counterparts; while \citet{wong2014robust} define robust versions of AIC, BIC and leave-one-out CV, all of them borrowing from the quasi-likelihood definition in \citet{cantoni2001}. The proposals based on brute-force (G)CV are generally too demanding to be practical for medium to large applications. The robust information criteria are more tractable, although still with a high computational cost if grid searches are to be used. In all of the three existing approaches, there is no formal treatment of the robustness tuning constant selection. \citet[p.~723]{alimadad2011outlier} advise to use $c=1.5$, commenting on the fact that ``values of $c$ between 1 and 4 produce similar qualitative results''. \citet[p.~33]{croux2012robust} suggest using $c=1.345$ for both estimating equations for the mean and the dispersion, borrowing from the Gaussian regression setting and stating that ``this value gives reasonable results for other models as well''. Finally, \citet{wong2014robust} suggest to use $c=1.6$ as in \citet{cantoni2001} without further discussion, even though the simulation designs are different.

Since we only have one smooth term here, we can afford the computational cost of the brute-force CV of AS and consider three variants of our estimator to compare smoothing parameter selection methods: minimizing our proposed robust AIC (RAIC); minimizing our robust BIC (RBIC); and the extended Fellner-Schall method (EFS). The RAIC/RBIC minimization are performed by a grid search with a relative numerical tolerance of $10^{-5}$ on the scale of the single smoothing parameter $\lambda$. All the methods have been tuned to achieve an MDP of 0.95 following the procedure introduced in Section~\ref{ss:choicetuning}, to make them comparable. The resulting tuning constants are $k=1.2$ for the AS method, \texttt{tccM} = 1.2 and \texttt{tccG} = 1.345 for CGP, $c=1.2$ for WYL, and $c=5.8$ for our approach. As already noted by \citet[][p.~286]{wong2014robust}, the CGP method estimates an additional dispersion parameter by default. This implies greater modeling flexibility and may make the comparison unfair in some situations, but we do not expect this to contribute much to its performance in the simulation settings considered here. We evaluate and compare the performances of the methods by assessing their MSE for the Poisson mean parameter $\mu$ computed as $\text{MSE}(\hat{\mu},\mu) = \frac{1}{n} \sum_{i=1}^n (\hat{\mu}_i - \mu_i)^2$. The \texttt{R} code for the WYL approach is available through the \texttt{R} package \texttt{robustGAM}, whereas the AS approach is available via the \texttt{R} package \texttt{rgam}. The code for the CGP approach was retrieved from the online supplementary material of \citet{wong2014robust}.

\begin{figure}
\centering
\includegraphics[width=0.8\textwidth]{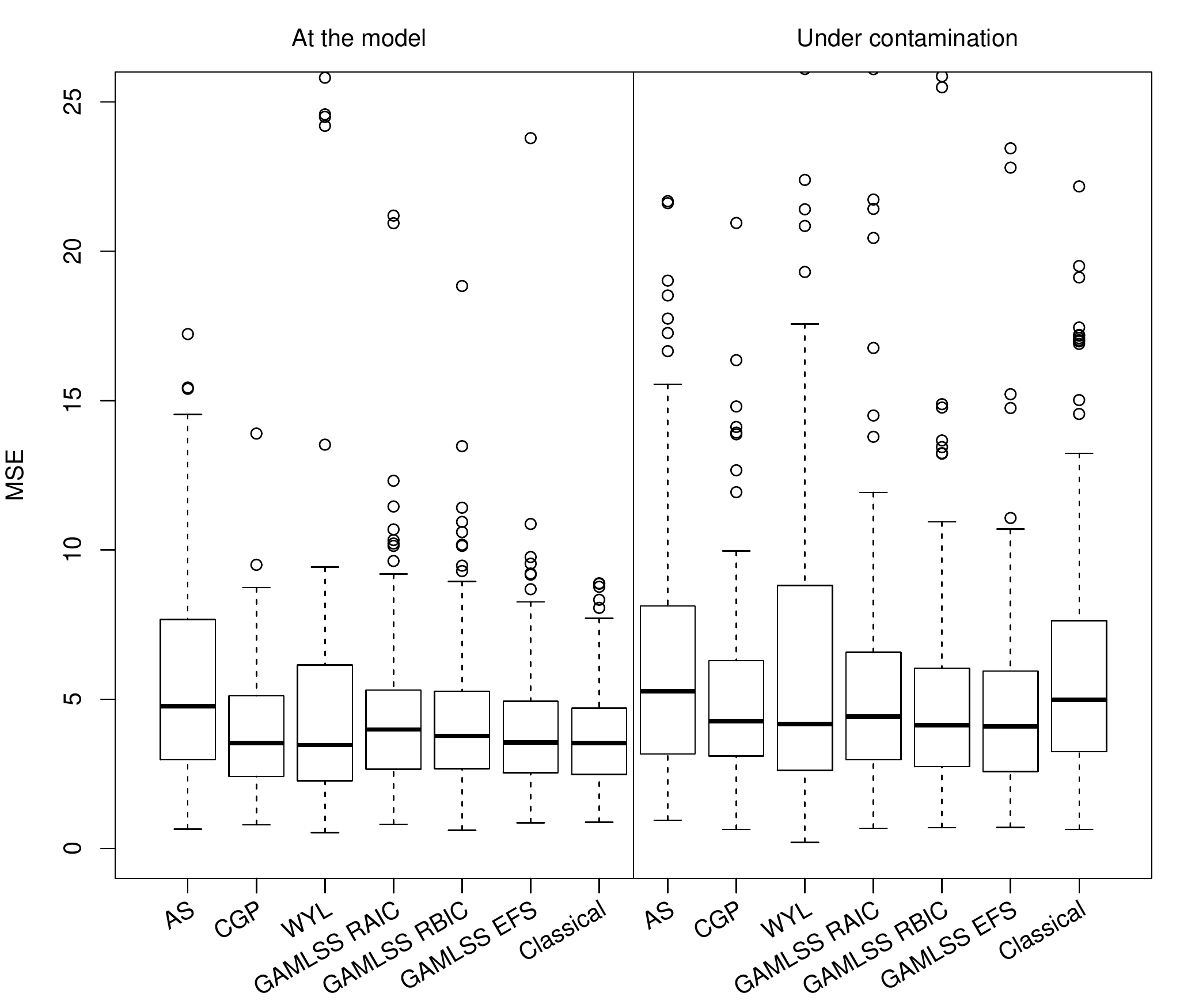}
\caption{GAM simulation, MSE at the assumed model and under contaminated data (vertical scale manually set for better visualization, some points not displayed).}
\label{fig:GAMall}
\end{figure}

Figure~\ref{fig:GAMall} displays boxplots of the MSE for all methods both at the assumed Poisson GAM model (left sub-panel) and under contamination (right sub-panel). Some numerical summary statistics are given in Table~\ref{tab:MSEallrobust}. The classical (ML-based) estimation has the lowest MSE under clean data, while it unsurprisingly shows poor performance under contamination. Among the robust methods, AS has the largest MSEs and tends to vary more than the others. CGP, WYL and our method all roughly have the same MSEs on average, although WYL shows larger variability across samples. Among our three variants (RAIC, RBIC and EFS) performance is similar at the model, but under contamination RAIC features slightly larger MSEs. This is in line with remarks made by \citet{wong2014robust} about AIC/RAIC favoring wigglier fits which here may allow contaminated observations to contribute relatively more to the fit than with heavier penalties such as BIC/RBIC, and this regardless of the robustness property of the method.

Overall, these simulation results yields two main conclusions. First, our proposed robust method performs similarly to the best-performing existing alternatives in the GAM special case. Second, the extended Fellner-Schall method allows for a reliable selection of the smoothing parameter and is on par with minimizing the RBIC but at a fraction of the computational cost of a grid search.

\begin{table}
\caption{GAM simulation, summary statistics of MSE for the robust methods (SD is standard deviation, IQR is inter-quartile range).}
\label{tab:MSEallrobust}
\centering
\begin{tabular}{lrrrrrr}
\hline
  & AS & CGP & WYL & GAMLSS RAIC & GAMLSS RBIC & GAMLSS EFS\\
\hline
\multicolumn{7}{c}{At the model}\\
  Average & 102.51 & 3.93 & 12.76 & 4.52 & 4.61 & 4.02 \\ 
  SD      & 282.49 & 1.98 & 24.00 & 2.79 & 3.70 & 2.37 \\ 
  Median  &   4.76 & 3.53 &  3.46 & 3.99 & 3.77 & 3.55 \\ 
  IQR     &   4.69 & 2.69 &  3.87 & 2.63 & 2.58 & 2.39 \\[1ex]
\multicolumn{7}{c}{Under contamination}\\
  Average &  61.40 &  7.31 & 16.39 &  2.50$^\dagger$ &  4.47$^\dagger$ & 20.20 \\ 
  SD      & 217.46 & 30.22 & 29.51 & 35.29$^\dagger$ & 63.25$^\dagger$ & 89.43 \\ 
  Median  &   5.27 &  4.27 &  4.17 &  4.41           &  4.13           & 4.09 \\ 
  IQR     &   4.94 &  3.15 &  6.17 &  3.59           &  3.26           & 3.35 \\ 
\hline
\multicolumn{7}{l}{$^\dagger${\small $\times 10^{14}$ due to two samples creating divergence}}\\
\end{tabular}
\end{table}

\section{Application to Brain Imaging Data}
\label{s:data}

The data motivating the proposed method come from fMRI of the human brain. These data were presented in \cite{Landau2003} and subsequently used in \cite{Wood17}, and are available in the \texttt{R} package \texttt{gamair} available on CRAN. The goal of the original study is to test for a difference in the timing (phase shift) of the physiological response between two anatomically distinct brain regions. For this purpose, a set of fMRI measures were acquired from a healthy participant during the performance of a verbal fluency task. The active task of this experiment consisted of generating words beginning with a cued letter, while the baseline condition was given by covertly repeating a letter. Brain activity was then summarized as the median of three measurements of fundamental power quotient on each brain voxel. The coordinates of each voxel were also recorded; these can be used to model the response surface. \citet[p.~329]{Wood17} identified two extreme voxel responses that were discarded for the subsequent analysis.

The response variable is thus the median fundamental power quotient \texttt{medFPQ} which represents the physiological response of the brain to controlled stimuli. This response is measured at voxels in a 2D brain slice with two covariates \texttt{X} and \texttt{Y} identifying the location of each voxel. The \texttt{medFPQ} measurements are rather noisy with possible spikes and troughs in activity which do not relate to the controlled stimulus, but the mean response level and its spread are likely to vary smoothly over the brain slice. Following \cite{Wood17}, we thus model both the mean and variance of \texttt{medFPQ} as joint functions \texttt{s(X,Y)} to be approximated by thin plate regression spline basis functions with a smoothness penalty based on second order derivatives. However, contrary to the analysis in \cite[][p.~329]{Wood17} where two voxels with \texttt{medFPQ} $\leq 5\times 10^{-3}$ were excluded on the ground that they can be regarded as outliers, we will consider the entire data set without exclusions. Given the nonnegative and positively skewed nature of \texttt{medFPQ}, we postulate a gamma distribution parameterized with mean $\mu$ and variance $\sigma^2\mu^2$, with $\log(\mu) = \eta_1 = s_1(\texttt{X},\texttt{Y})$ and $\log(\sigma) = \eta_2 = s_2(\texttt{X},\texttt{Y})$. Other families were considered, including the log-logistic distribution which is outside the exponential family; diagnostics and model validation (not presented here) confirmed that a gamma distribution provides the best fit.

We fit the gamma GAMLSS with a classical (ML, non-robust) estimation method and our proposed robust method. Because of the joint smoother used here, we rely on the EFS method which provides fast computations. The robust estimator is tuned to achieve an MDP of 0.95, resulting in a robustness constant of $c=4.5$.

\begin{figure}
\centering
\includegraphics[width=1\textwidth]{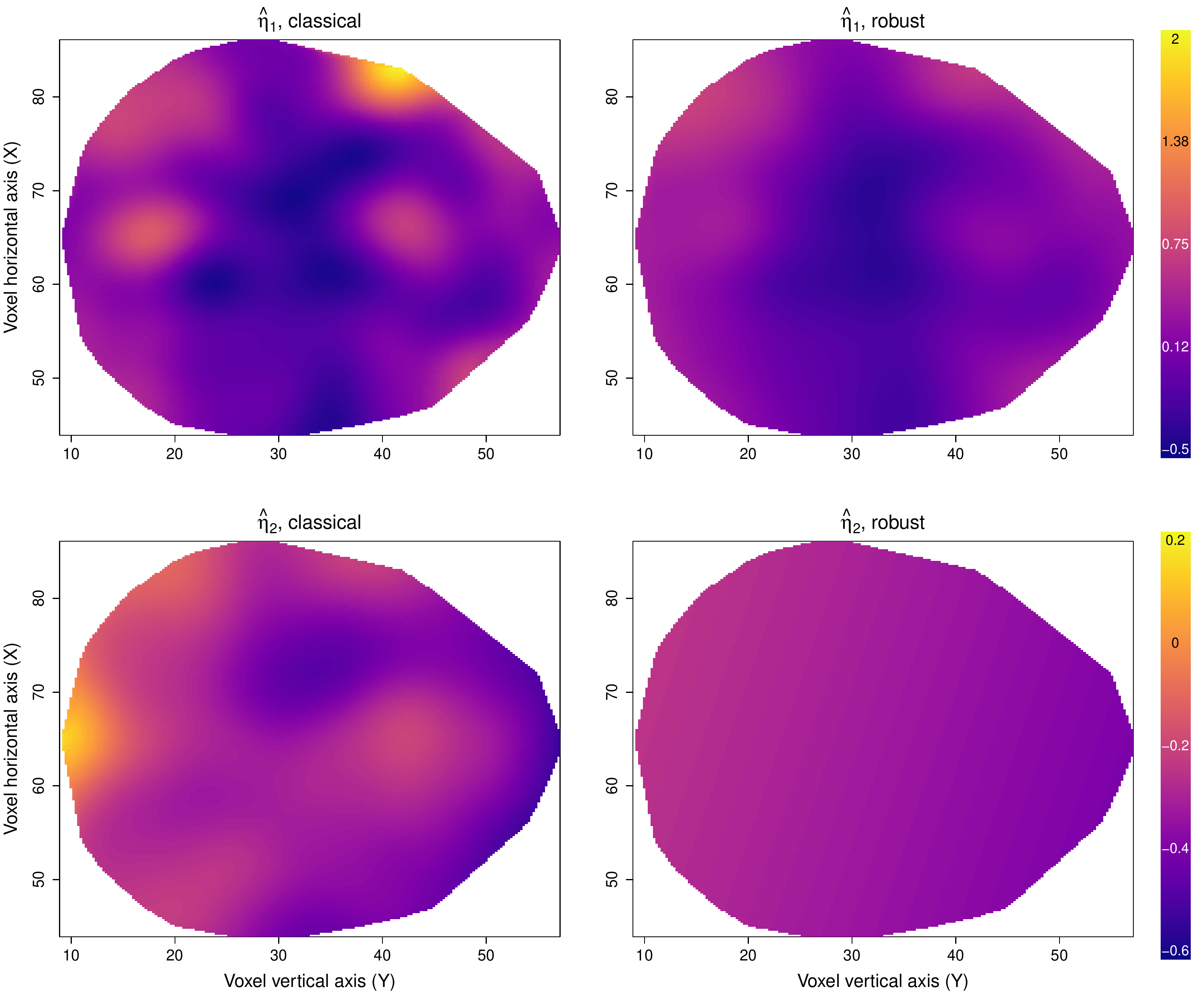} 
\caption{Brain imaging data, fitted surfaces for the linear predictors $\eta_1$ (top row) and $\eta_2$ (bottom row) based on classical (left column) and robust (right column) estimation methods.}
\label{fig:BrainEta1Eta2}
\end{figure}

The fitted surfaces for $\eta_1$ and $\eta_2$ are given in Figure~\ref{fig:BrainEta1Eta2}. Overall, the robust fitted surfaces appear smoother for both parameters, with a surface that is nearly flat for $\eta_2$. The classical fit uses a total of 77.2 effective degrees of freedom (56.09 for fitting $\eta_1$, 19.11 for $\eta_2$, plus 2 for the constants), whereas the robust fit only uses 30.04 effective degrees of freedom (26.00 for fitting $\eta_1$, 2.04 for $\eta_2$, and 2 for the constants). This hints that the automatic selection of the smoothing parameter in the classical fit was influenced by some potentially outlying observations.

\begin{figure}
\centering
\includegraphics[width=0.75\textwidth]{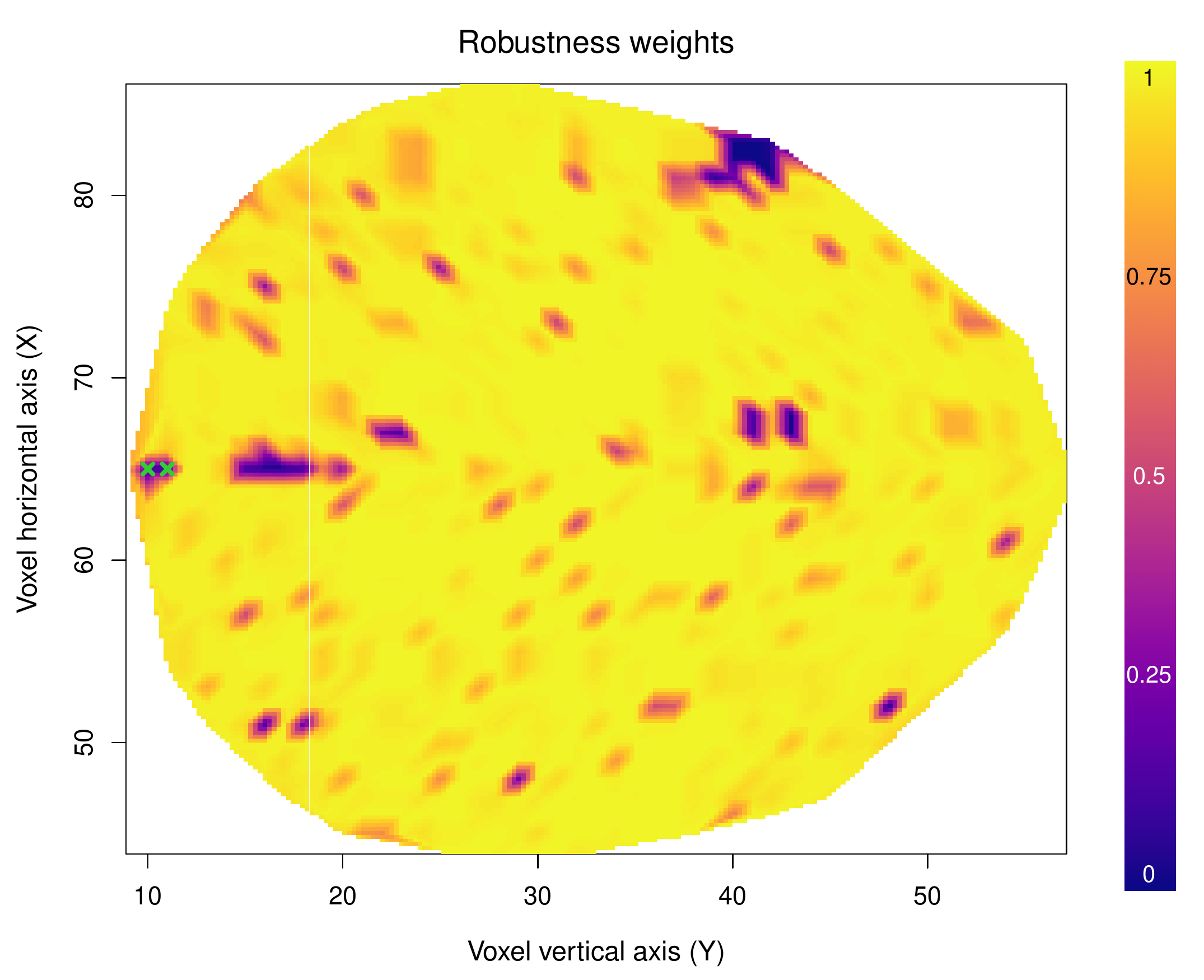}
\caption{Brain imaging data, robustness weights from the robust GAMLSS fit, with the two green crosses identifying the two observations excluded from the analysis in \citet{Wood17}.}
\label{fig:Brain_m3rob_weight}
\end{figure}

Consider the largest local differences in Figure~\ref{fig:BrainEta1Eta2} between the two fits: in the upper-right corner of the brain for $\hat{\eta}_1$ (which has motivated the contamination scheme of Section~\ref{ss:simgamlss}), and in the leftmost part of the brain for $\hat{\eta}_2$. For the latter, classical estimates imply a much larger localized response variance than robust estimates do. This is driven by two observations in this area which are the ones excluded from the analysis in \cite{Wood17}. But for the large difference in $\hat{\eta}_1$, the spike in mean brain activity implied by classical estimates is much subdued when considering robust estimation. This is explained when investigating the robustness weights, which are displayed in Figure~\ref{fig:Brain_m3rob_weight}. A few observations in the top-right corner are heavily downweighted by the robust method, which results in the smoother mean surface in Figure~\ref{fig:BrainEta1Eta2}. These low weights do not imply that these observations are necessarily outliers, but simply that they do not seem to follow the same trends as the majority of the data given the gamma GAMLSS assumed here. The downweighted observations in the top-right corner may indeed represent a physiological response of interest here, we note that the robustness weights identify them in an automated way. Also, note that the two observations excluded by \cite{Wood17} are also heavily downweighted; these are indicated in Figure~\ref{fig:Brain_m3rob_weight} as green crosses for reference. Hence, the robust fitted surfaces combined with the robustness weights are effective at both modeling smooth functions in a reliable way and at automatically detecting observations deviating from trends and model assumptions.

\section{Discussion}
\label{s:discu}

We introduced a robust estimation method for the broad class of GAMLSS. Our approach is quite general since it can be employed for any differentiable likelihood. By directly robustifying the log-likelihood and correcting it for Fisher consistency, this method yields natural robust versions of AIC and BIC. For more complicated designs where grid searches are not feasible, our extended Fellner-Schall method allows for a reliable and automatic selection of smoothing parameters. Our implementation in the \texttt{R} package \texttt{GJRM}, based on the trust region algorithm, is modular and stable. Furthermore, the introduced MDP addresses the challenge of the selection of the robustness tuning constant for models with flexible non-linear effects in a simple and effective way. We believe this criterion has broad applicability in the implementation of robust methods in many contexts, including the ones where efficiency criteria based on asymptotic covariances already exist but may be computationally expensive.

Simulations in the special case of a GAM showed that our robust estimator is on par with the best-performing existing approaches, when tuned for comparable robustness under the assumed model. Simulations in the broader GAMLSS setting as well as our application to the brain imaging data showed that our robust estimator allows for the automatic detection of deviating observations through the robustness weights, and that the approach yields trustworthy estimates.


Future work includes the extension to high-dimensional settings, following for instance \cite{mayr2012generalized} where the problem of variable selection is considered. An alternative strategy for variable selection is developed in \cite{hambuckers2018understanding} and \cite{RePEc:inn:wpaper:2018-16} using $L_1$-type of penalties.









\renewcommand{\thesection}{\Alph{section}}
\setcounter{equation}{0}
\setcounter{section}{0}

\pagebreak

\begin{center}
\Large{Web Appendices for \textit{Robust Fitting for Generalized Additive Models for Location, Scale and Shape}}

\bigskip
\normalsize{William H.\ Aeberhard$^{1}$, Eva Cantoni$^{2}$,\\Giampiero Marra$^{3}$, and Rosalba Radice$^{4}$}\\[0.7ex]
\footnotesize{$^{1}$Department of Mathematical Sciences, Stevens Institute of Technology, USA}\\[0.7ex]
\footnotesize$^{2}$Research Center for Statistics and GSEM, University of Geneva, Switzerland\\[0.7ex]
\footnotesize$^{3}$Department of Statistical Science, University College London, UK\\[0.7ex]
\footnotesize$^{4}$Cass Business School, City, University of London, UK\\
\end{center}


\section*{Web Appendix~A: Conditions and Proof of Theorem~1}

The following conditions are required for Theorem~1 in the main body of the paper.
\begin{itemize}
\item[(C1)]{
For any $c < \infty$, $\rho_c:\mbbR\rightarrow\mbbR$ is convex, monotonically increasing, twice continuously differentiable, and has a first derivative $\rho_c'$ that is bounded within $[0,1]$.
}
\item[(C2)]{
The parameter space $\bDelta \subset \mbbR^p$ is compact and $\bdelta_0$ is an interior point of $\bDelta$. 
}
\item[(C3)]{
For any $y$ in the support, the mapping $\bdelta \rightarrow f(y|\mu,\sigma,\nu)$ is twice continuously differentiable with respect to $\bdelta$ and $\sup_{\bdelta \in \bDelta}f(y|\mu,\sigma,\nu) < \infty$.
}
\item[(C4)]{
For all $\bdelta$ in some open neighborhood around $\bdelta_0$, $\bM(\bdelta)$ is positive definite and $\bQ(\bdelta) < \infty$. 
}
\item[(C5)]{
The smoothness parameter $\blambda = o(n^{1/2})\bone$, i.e.\ a vanishing penalty $n^{-1}\blambda = o(n^{-1/2})\bone$ for the log-likelihood scaled by $1/n$. 
}
\end{itemize}


Condition~(C1) summarizes the requirements for the user-specified $\rho$ function. In particular, the boundedness of the first derivative is the key to the robustness property of the proposed estimator as it ensures that the influence function is itself bounded (see below). These requirements are satisfied by the log-logistic function used in the main body of the paper.


The conditions in (C2) essentially guarantee that the asymptotic distribution is absolutely continuous (i.e.\ without point-masses on the boundary of the domain). The compactness requirement is rather standard in non-parametric and sieve estimation contexts. 

Condition~(C3) includes most commonly used families of distributions. We note it may be relaxed to Lipschitz requirements on the derivative of $\log f(y|\mu,\sigma,\nu)$ and may hold only for asymptotically non-null sets \citep[see][]{huber1967}.

The requirements in (C4) are standard minimal conditions in the robustness literature, akin to assuming that the Fisher information matrix exists in a neighborhood of the true parameter in a maximum likelihood (ML) framework. 

Finally, (C5) imposes a minimum rate at which $\blambda$ decreases as $n\rightarrow\infty$. This rate is necessary to guarantee that the penalization does not induce any asymptotic bias for the robust estimator. 


\textit{Proof of Theorem~1.} The penalized robust estimator $\hat{\bdelta}$ is the solution in $\bdelta$ to
\begin{align} \label{eq:robesteq}
\mathbf{0} &= \frac{\partial \tilde{\ell}(\bdelta)}{\partial \bdelta} - \frac{1}{n}\bS(\blambda)\bdelta,
\end{align}
where here $\tilde{\ell}(\bdelta) =  \frac{1}{n}\sum_{i=1}^n \rho_c\big(\ell(\bdelta)_i\big) - \frac{1}{n}b_{\rho}(\bdelta)$ is the scaled robustified log-likelihood.
Given the smoothness of $\tilde{\ell}(\bdelta)$ ensured by (C1) and (C3), a first-order Taylor expansion about $\bdelta_0$ of the right-hand side of (\ref{eq:robesteq}) evaluated at the solution $\hat{\bdelta}$  yields
\begin{align*}
\mathbf{0} &= \left.\frac{\partial \tilde{\ell}(\bdelta)}{\partial \bdelta}\right|_{\bdelta=\bdelta_0} - n^{-1}\bS(\blambda)\bdelta_0 + \left(\left.\frac{\partial^2 \tilde{\ell}(\bdelta)}{\partial\bdelta\partial\bdelta\ts}\right|_{\bdelta=\bdelta_0} - n^{-1}\bS(\blambda)\right) (\hat{\bdelta}-\bdelta_0) + O_p(||\hat{\bdelta}-\bdelta_0||_2^2).
\end{align*}
Multiplying by $\sqrt{n}$ on both sides and rearranging lead to
\begin{align}\label{eq:}
\left(\left.\frac{\partial^2 \tilde{\ell}(\bdelta)}{\partial\bdelta\partial\bdelta\ts}\right|_{\bdelta=\bdelta_0} - n^{-1}\bS(\blambda)\right)&\sqrt{n}(\hat{\bdelta}-\bdelta_0)\nonumber\\
&= -\sqrt{n} \left.\frac{\partial \tilde{\ell}(\bdelta)}{\partial \bdelta}\right|_{\bdelta=\bdelta_0} + n^{-1/2}\bS(\blambda)\bdelta_0 + O_p(\sqrt{n}||\hat{\bdelta}-\bdelta_0||_2^2).
\end{align}
Considering the limit as $n \rightarrow \infty$, (C5) implies $n^{-1}\bS(\blambda) \rightarrow \mathbf{0}$ and $n^{-1/2}\bS(\blambda)\bdelta_0 \rightarrow \mathbf{0}$. This effectively reduces the estimating equations to those of the unpenalized robust estimator that solves $\partial \tilde{\ell}(\bdelta)/\partial \bdelta = \mathbf{0}$. Viewing $\bdelta_0$ as the true parameter that generated the data, we can thus invoke standard results from $M$-estimation theory. In particular, conditions (C1)--(C4) are sufficient to apply Theorem~2 of \citet{huber1967} so that the unpenalized robust estimator converges a.s.\ to $\bdelta_0$. This implies that the statistical functional $\bT(F)$ defined as the solution in $\bdelta$ to $\E_F\left[\psi(Y,\bdelta)\right] = \mathbf{0}$, where
\begin{equation*}
\psi(Y,\bdelta) = \rho_c'\big(\log f(Y|\mu,\sigma,\nu)\big)\frac{\partial \log f(Y|\mu,\sigma,\nu)}{\partial \bdelta} - \frac{\partial }{\partial \bdelta}\E\left[ \rho_c^\star\big(\log f(Y|\mu,\sigma,\nu)\big) \right],
\end{equation*}
satisfies $\bT(D) = \bdelta_0$, i.e.\ it is Fisher consistent for $\bdelta_0$. Asymptotic normality follows under (C1)--(C4) by observing that the remainder in (\ref{eq:}) is $O_p(n^{-1/2})$ because of $\sqrt{n}$-consistency and by Theorem~3 of \citet{huber1967}:
\begin{equation*}
\sqrt{n}(\hat{\bdelta}-\bdelta_0) \underset{n\rightarrow\infty}{\overset{D}{\longrightarrow}} \text{N}(\mathbf{0}, \bV(\bdelta_0))
\end{equation*}
where $\bV(\bdelta) = \bM(\bdelta)^{-1} \bQ(\bdelta) \bM(\bdelta)^{-\sf T}$.

\section*{Web Appendix~B: Extended Fellner-Schall Method}

We follow here the main steps of \citet{wood2017fellnerschall} in developing a robust extended Fellner-Schall (EFS) method to select the smoothness parameter $\blambda$.

The quadratic penalty in the penalized robustified log-likelihood
\begin{equation*}
\tilde{\ell}_{p}(\bdelta) = \tilde{\ell}(\bdelta) - \frac{1}{2} \bdelta\ts \bS \bdelta
\end{equation*}
can be viewed as an improper Gaussian prior distribution on $\bdelta$, seen as a random vector throughout this section, with $\blambda$ now seen as another model parameter. For this, consider adding the term corresponding to the determinant of the Gaussian covariance matrix to get an improper prior, i.e.\ up to a constant term (involving $\sqrt{2\pi}$). We thus define the joint robustified log-likelihood $l$ for $\bY=(Y_1,\ldots,Y_n)$ and $\bdelta$ as
\begin{align*}
l(\by,\bdelta;\blambda) &= \tilde{\ell}_p(\bdelta) - \frac{1}{2} \log |\bS(\blambda)^{-1}| = \tilde{\ell}(\bdelta) - \frac{1}{2} \bdelta^\top \bS(\blambda) \bdelta + \frac{1}{2} \log |\bS(\blambda)|
\end{align*}
where $|\cdot|$ denotes matrix determinant, so that we can view $L(\by,\bdelta;\blambda) = \exp(l(\by,\bdelta))$ as a joint (robustified) likelihood up to constant terms:
\begin{align}\label{eq:jointroblkhd}
L(\by,\bdelta;\blambda) &= \exp\big(\tilde{\ell}(\bdelta)\big) \times \exp\left(-\frac{1}{2} \bdelta^\top \bS(\blambda) \bdelta\right)\frac{1}{\sqrt{|\bS(\blambda)^{-1}|}}.
\end{align}
The multivariate Gaussian distribution of $\bdelta$ has a mean vector of zero and covariance $\bS(\blambda)^{-1}$, or precision matrix $\bS(\blambda)$. By construction, $\bS(\blambda)$ is symmetric and positive definite so that all its eigenvalues are strictly positive and it is thus invertible. Note that this added term does not depend on $\bdelta$ so that our robust estimator $\hat{\bdelta}$ can also be defined as $\argmax_{\bdelta} l(\by,\bdelta;\blambda)$.

We derive the marginal robustified likelihood for $\bY$ by integrating out $\bdelta$ from $L(\by,\bdelta;\blambda)$: $L(\by;\blambda) = \int_{\bDelta} L(\by,\bdelta;\blambda) \,\td \bdelta$. The Laplace approximation of this integral can be summarized as a second-order Taylor expansion of $l(\by,\bdelta;\blambda)$ about the global maximum $\hat{\bdelta} = \hat{\bdelta}(\blambda) = \argmax_{\bdelta} l(\by,\bdelta;\blambda)$ for a given $\blambda$ and observation vector $\by$. This yields the following approximation to the marginal robustified log-likelihood
\begin{align}\label{eq:laplacemargloglkhd}
\log L(\by;\blambda) \approx l_{\text{LA}}(\by;\blambda) &= \tilde{\ell}(\hat{\bdelta}) - \frac{1}{2} \hat{\bdelta}^\top \bS(\blambda) \hat{\bdelta} + \frac{1}{2} \log |\bS(\blambda)| - \frac{1}{2}\log |\widehat{\bM}_p(\hat{\bdelta},\blambda)|,
\end{align}
where the ``LA'' subscript identifies Laplace-approximated quantities and the negative Hessian
\begin{equation*}
\widehat{\bM}_p(\bdelta,\blambda) = \frac{\partial^2 l(\by,\bdelta;\blambda)}{\partial \bdelta\partial\bdelta^\top} = - \frac{\partial^2 \tilde{\ell}(\bdelta)}{\partial \bdelta\partial\bdelta^\top} + \bS(\blambda) = \widehat{\bM}(\bdelta) + \bS(\blambda)
\end{equation*}
is guaranteed to be positive definite at least in a neighborhood of $\hat{\bdelta}$ for sufficiently large $n$ under the conditions of Theorem~1.

Now, viewing $\blambda$ as the parameter of this (Laplace-approximated) marginal log-likelihood, ideally we would then compute the first-order derivative:
\begin{align}\label{eq:derivLAmargrobloglkhd}
\frac{\partial l_{\text{LA}}(\by;\blambda)}{\partial \lambda_j} &= \frac{\partial \tilde{\ell}(\hat{\bdelta}(\blambda))}{\partial \lambda_j} - \frac{1}{2} \frac{\partial}{\partial \lambda_j} \hat{\bdelta}(\blambda)^\top \bS(\blambda) \hat{\bdelta}(\blambda) + \frac{1}{2} \frac{\partial}{\partial \lambda_j} \log |\bS(\blambda)| - \frac{1}{2} \frac{\partial}{\partial \lambda_j} \log |\widehat{\bM}_p(\hat{\bdelta},\blambda)|.
\end{align}
Unfortunately, it cannot be computed directly in general since $\blambda$ appears essentially everywhere. In particular, $\blambda$ appears in $\widehat{\bM}_p$ both explicitly because $\widehat{\bM}_p$ is a direct function of $\bS(\blambda)$ (see above) and implicitly through $\hat{\bdelta}(\blambda)$. We therefore need to neglect the dependence on $\blambda$ for some terms in (\ref{eq:derivLAmargrobloglkhd}). Specifically, we are going to consider $\hat{\bdelta}$ as a function of a previous ``inactive'' iterate $\blambda'$, so that $\hat{\bdelta}$ is fixed with respect to the current ``active'' $\blambda$. This implies that $\widehat{\bM}_p(\hat{\bdelta},\blambda) = - \left.\frac{\partial^2 \tilde{\ell}(\bdelta)}{\partial \bdelta\partial\bdelta^\top}\right|_{\bdelta=\hat{\bdelta}} + \bS(\blambda)$ is now a function of the active $\blambda$ only explicitly through $\bS(\blambda)$. That way, the new Laplace-approximated marginal log-likelihood, written as $l_{\text{LA}}(\by;\blambda,\blambda')$, is now a function of the active $\blambda$ only through $\bS(\blambda)$. To compute the derivative of $l_{\text{LA}}(\by;\blambda,\blambda')$ with respect to the active $\blambda$, we adapt Jacobi's formula to invertible matrices and obtain
\begin{equation*}
\frac{\partial \log|\bS(\blambda)|}{\partial \lambda_j} = \tr\left\{\bS(\blambda)^{-1}\frac{\partial \bS(\blambda)}{\partial \lambda_j}\right\}
\end{equation*}
since $\bS(\blambda)$ is symmetric and invertible (and a differentiable function of $\blambda$). Based on this result, we can get a simplified version of (\ref{eq:derivLAmargrobloglkhd}):
\begin{align}\label{eq:derivLAmargrobloglkhdactive}
\frac{\partial l_{\text{LA}}(\by;\blambda,\blambda')}{\partial \lambda_j} &=
- \frac{1}{2} \hat{\bdelta}^\top \frac{\partial \bS(\blambda)}{\partial \lambda_j} \hat{\bdelta}
+ \frac{1}{2} \tr\left\{\bS(\blambda)^{-1}\frac{\partial \bS(\blambda)}{\partial \lambda_j}\right\}
- \frac{1}{2} \tr\left\{\widehat{\bM}_p(\hat{\bdelta},\blambda)^{-1}\frac{\partial \bS(\blambda)}{\partial \lambda_j}\right\}.
\end{align}
In this new expression, $\partial \bS(\blambda)/\partial \lambda_j$ is straightforward to write down and implement since $\bS(\blambda)$ is block-diagonal and each block is linear in the components of $\blambda$ and only involves the (known) basis functions.

Now we can follow the same heuristic reasoning as in Section~2 of \citet{wood2017fellnerschall}. There are four necessary requirements for the update of $\lambda_j$:\begin{enumerate}
\item{$\lambda_j$ must not change if $\partial l_{\text{LA}}(\by;\blambda,\blambda')/\partial \lambda_j = 0$.
}
\item{$\lambda_j$ needs to be decreased (because we maximize $l_{\text{LA}}$) if $\partial l_{\text{LA}}(\by;\blambda,\blambda')/\partial \lambda_j < 0$, this happens iff
\begin{equation*}
\tr\big\{\bS(\blambda)^{-1} \partial \bS(\blambda)/\partial \lambda_j\big\} - \tr\big\{\widehat{\bM}_p(\hat{\bdelta},\blambda)^{-1} \partial \bS(\blambda)/\partial \lambda_j \big\} < \hat{\bdelta}^\top \big(\partial \bS(\blambda)/\partial \lambda_j\big) \hat{\bdelta}.
\end{equation*}
}
\item{
Similarly, $\lambda_j$ needs to be increased if $\partial l_{\text{LA}}(\by;\blambda,\blambda')/\partial \lambda_j > 0$, this happens iff
\begin{equation*}
\tr\big\{\bS(\blambda)^{-1} \partial \bS(\blambda)/\partial \lambda_j\big\} - \tr\big\{\widehat{\bM}_p(\hat{\bdelta},\blambda)^{-1} \partial \bS(\blambda)/\partial \lambda_j \big\} > \hat{\bdelta}^\top \big(\partial \bS(\blambda)/\partial \lambda_j\big) \hat{\bdelta}.
\end{equation*}
}
\item{
$\lambda_j$ must remain positive.
}
\end{enumerate}
Before moving on, we need to check four conditions to apply Theorem~1 of \citet{wood2017fellnerschall}: (i) the negative Hessian $- \frac{\partial^2 \tilde{\ell}(\bdelta)}{\partial \bdelta\partial\bdelta^\top}$ is positive definite; (ii) $\bS(\blambda)$ is positive semi-definite; (iii) the null space of $\bS(\blambda)$ is independent of $\blambda$; (iv) $\partial \bS(\blambda)/\partial \lambda_j$ is positive semi-definite for all $j$. Condition (i) is satisfied under the conditions of our Theorem~1 in the main body of the paper, at least in a neighborhood of $\hat{\bdelta}$ for sufficiently large $n$ based on some reasonable previous iterate $\blambda'$. Condition (ii) directly follows from our construction of $\bS(\blambda)$ as a positive definite matrix. Condition (iii) holds because $\bS(\blambda)$ is invertible and thus its null space is $\{\mathbf{0}\}$. Finally, for Condition~(iv), recall that $\bS(\blambda)$ is typically block-diagonal and that its blocks (themselves block-diagonal) are usually linear in all elements of $\blambda$. The coefficients typically are integrated quadratic forms of the basis functions (we penalize the wiggliness of the fitted smooth function), so the derivative $\partial \bS(\blambda)/\partial \lambda_j$ typically is positive definite so that Condition~(iv) is satisfied. Hence Theorem~1 in \citet{wood2017fellnerschall} holds: $\tr\big\{\bS(\blambda)^{-1} \partial \bS(\blambda)/\partial \lambda_j\big\} - \tr\big\{\widehat{\bM}_p(\hat{\bdelta},\blambda)^{-1} \partial \bS(\blambda)/\partial \lambda_j \big\} > 0$. Further, by the same arguments for checking Condition~(iv), the quadratic form $\hat{\bdelta}^\top \big(\partial \bS(\blambda)/\partial \lambda_j\big) \hat{\bdelta}$ is always positive.

This allows us to consider an analogous expression to that of Equation~(8) in \citet{wood2017fellnerschall} to update all elements of $\blambda$ from iteration $[k]$ to $[k+1]$:
\begin{align}
\lambda^{[k+1]}_j &= \lambda^{[k]}_j \times \frac{\tr\big\{\bS(\blambda^{[k]})^{-1} \left.\partial \bS(\blambda)/\partial \lambda_j\right|_{\blambda=\blambda^{[k]}}\big\} - \tr\big\{\widehat{\bM}_p(\hat{\bdelta},\blambda^{[k]})^{-1} \left.\partial \bS(\blambda)/\partial \lambda_j\right|_{\blambda=\blambda^{[k]}} \big\}}{\hat{\bdelta}^\top \big(\partial \bS(\blambda)/\partial \lambda_j|_{\blambda=\blambda^{[k]}}\big) \hat{\bdelta}},
\end{align}
where $\hat{\bdelta} = \hat{\bdelta}(\blambda^{[k]})$ here. This heuristic update satisfies all four requirements above since the ratio multiplying $\lambda_j^{[k]}$ is always positive and equals one if $\partial l_{\text{LA}}(\by;\blambda,\blambda')/\partial \lambda_j = 0$. As noted by \citet[][p.~1073]{wood2017fellnerschall}, this ratio is not guaranteed to have the optimal step size, e.g.\ we could under-shoot the optimal update in terms of increasing $l_{\text{LA}}(\by;\blambda,\blambda')$. So a line search at each update may be useful, although the additional computational cost might offset the gain in reducing the number of iterations. 

\pagebreak
\section*{Web Appendix~C: Supplementary Figures}

\renewcommand{\thefigure}{S\arabic{figure}} 
\setcounter{figure}{0}                      


\begin{figure}[!ht]
\centering
\includegraphics[width=1\textwidth]{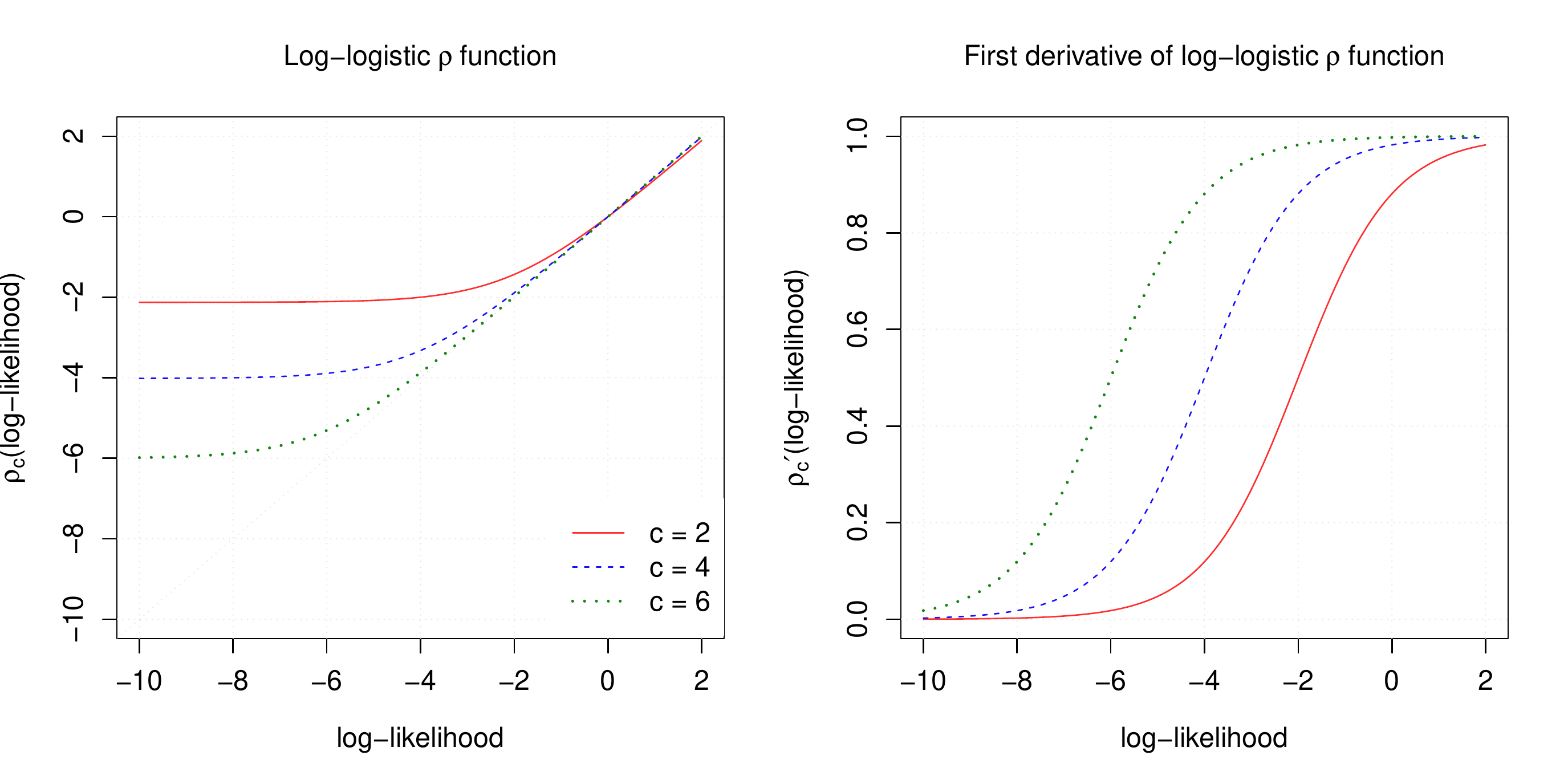}
\caption{Log-logistic $\rho_c$ function (left panel) and its first derivative $\rho_c'$ (right panel), for some values of the robustness tuning constant $c$.}
\label{fig:rho}
\end{figure}

\begin{figure}[!ht]
\centering
\includegraphics[width=1\textwidth]{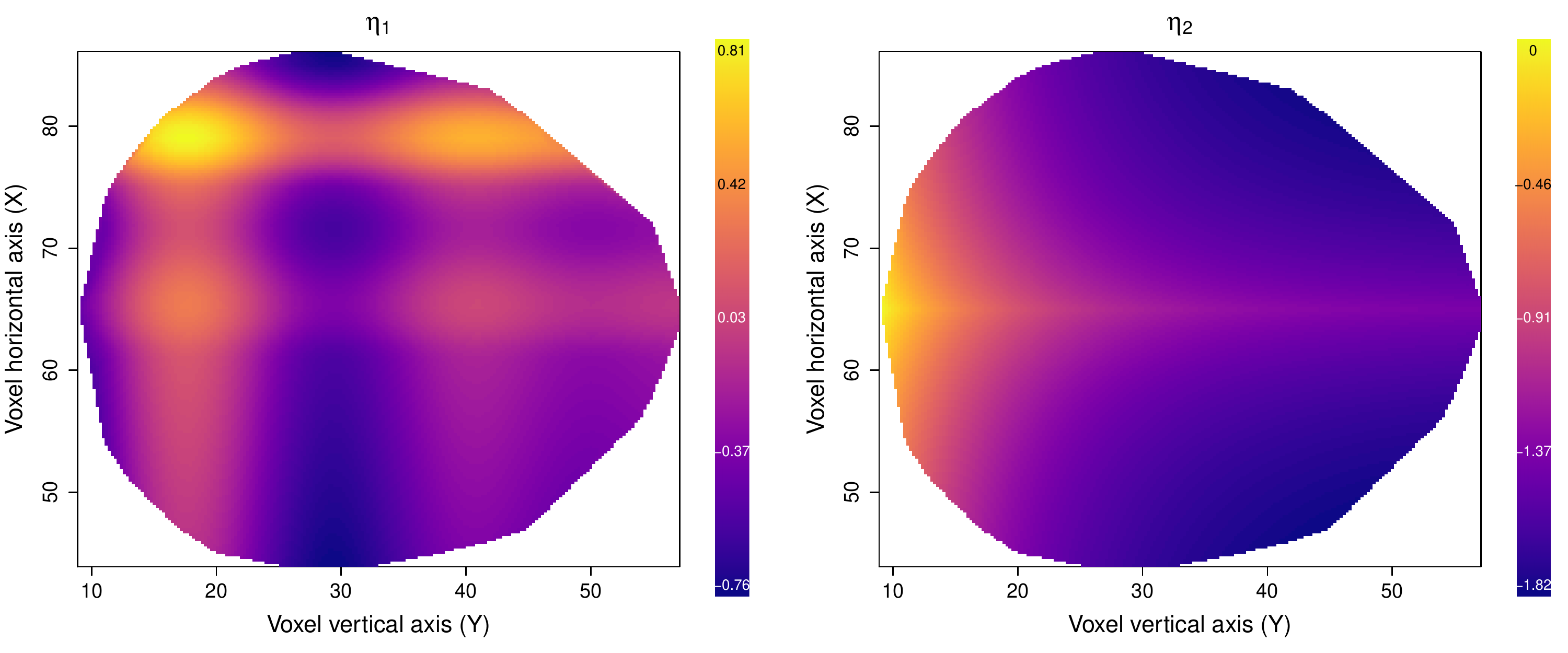}
\caption{GAMLSS simulation, surfaces of $s_1$ and $s_2$ functions used to simulate data.}
\label{fig:simgamlssdesign}
\end{figure}

\begin{figure}[!ht]
\centering
\includegraphics[width=1\textwidth]{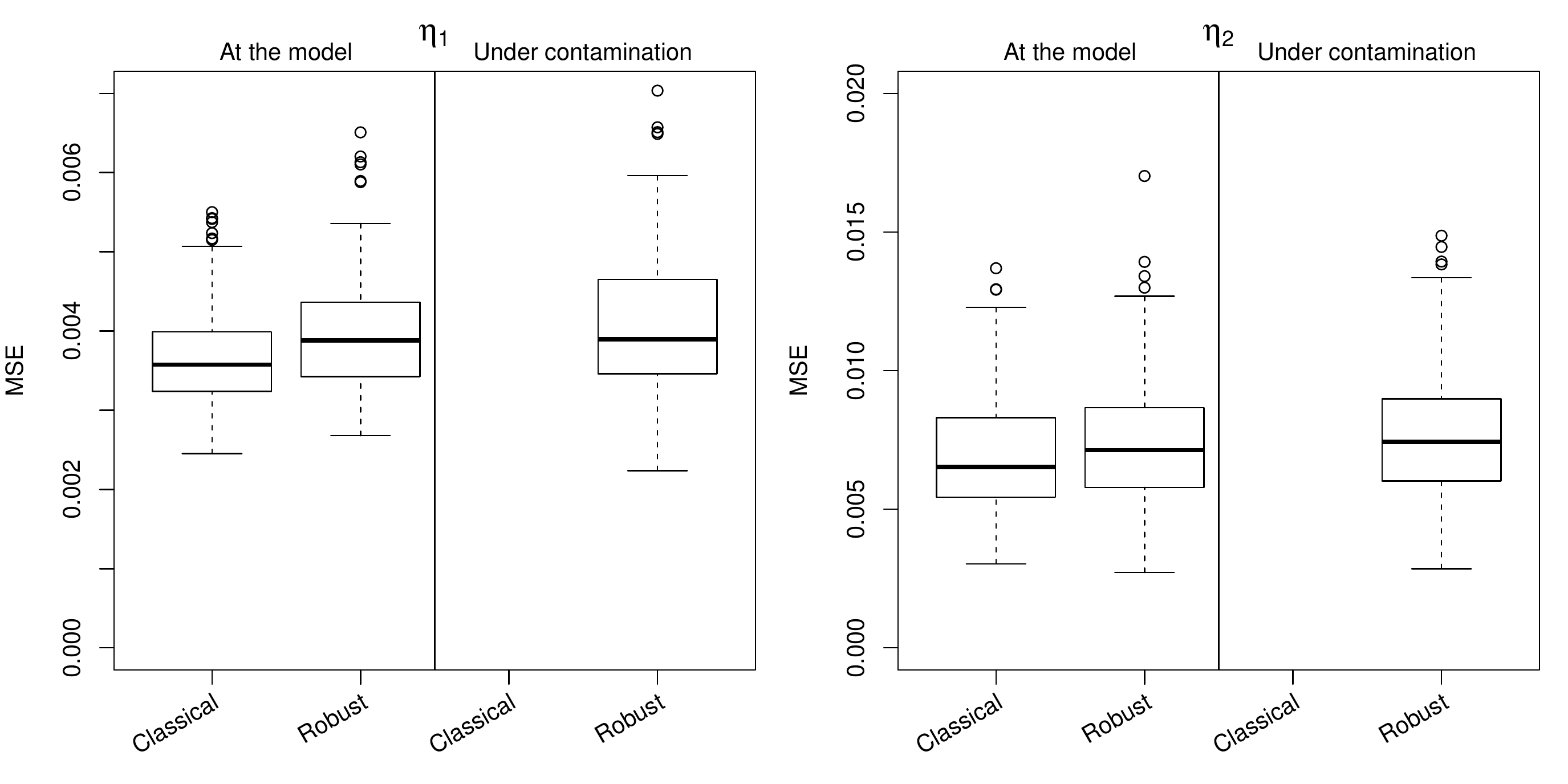}
\caption{GAMLSS simulation, MSE for the linear predictors $\eta_1$ (left panel) and $\eta_2$ (right panel) for classical and robust methods with data generated at the assumed model and under contamination. Vertical scales set manually for better visualization.}
\label{fig:simgamlss.mse.etazoomed}
\end{figure}

\begin{figure}[!ht]
\centering
\includegraphics[width=1\textwidth]{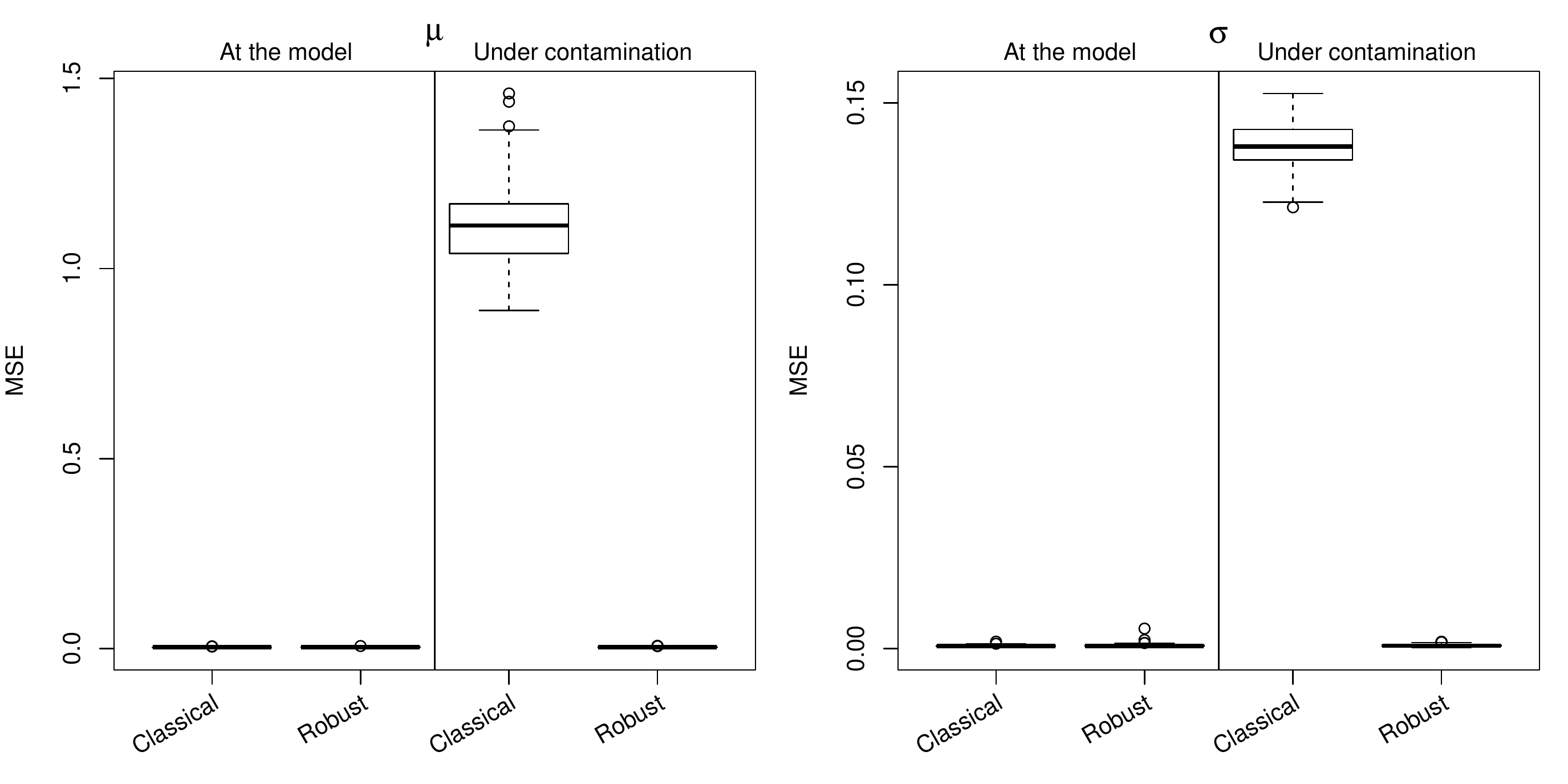}
\caption{GAMLSS simulation, MSE for the canonical parameters $\mu$ (left panel) and $\sigma$ (right panel) for classical and robust methods with data generated at the assumed model and under contamination.}
\label{fig:simgamlss.mse.musigma}
\end{figure}

\begin{figure}[!ht]
\centering
\includegraphics[width=1\textwidth]{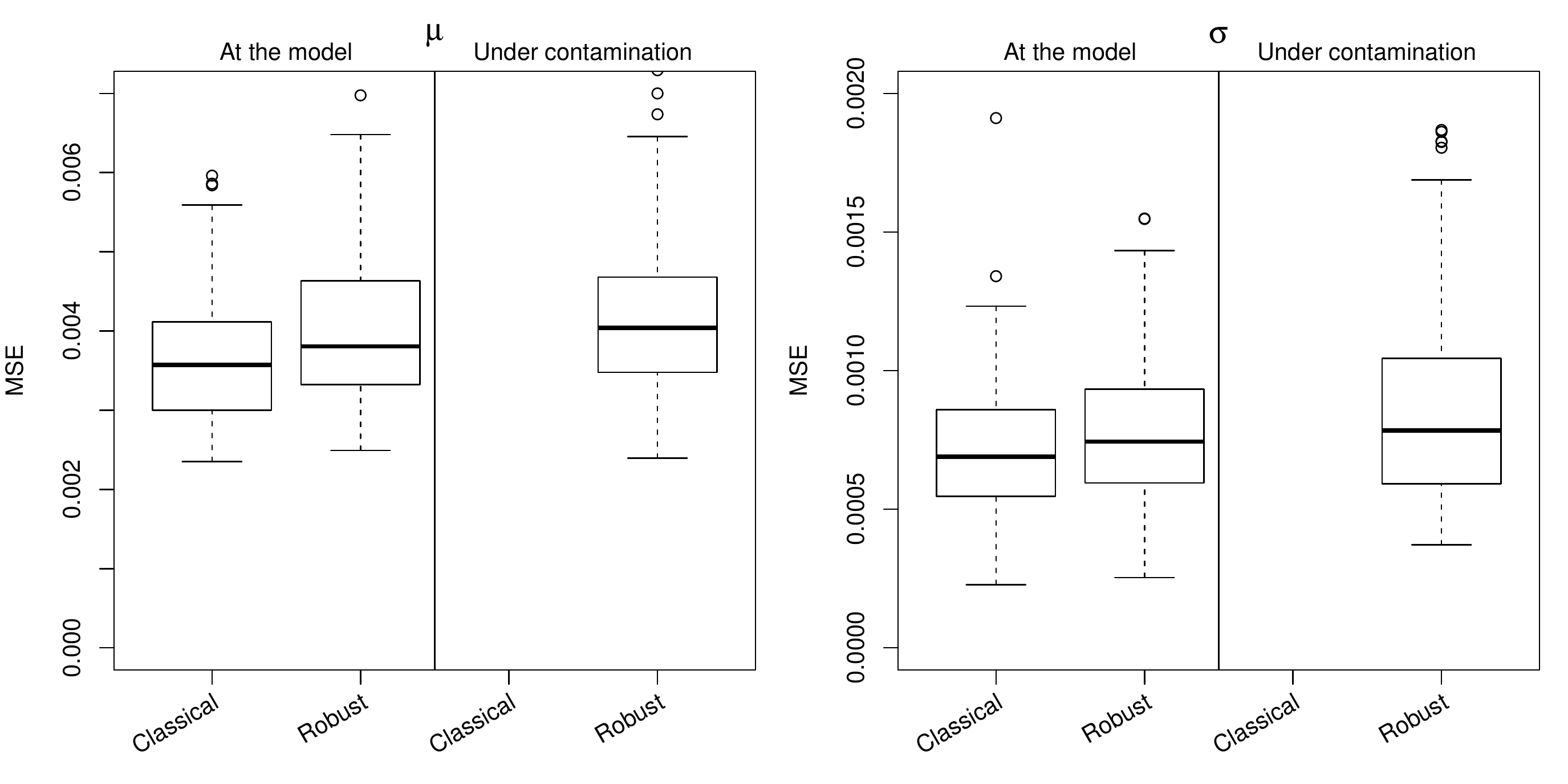}
\caption{GAMLSS simulation, MSE for the canonical parameters $\mu$ (left panel) and $\sigma$ (right panel) for classical and robust methods with data generated at the assumed model and under contamination. Vertical scales set manually for better visualization.}
\label{fig:simgamlss.mse.musigmazoomed}
\end{figure}

\begin{figure}[!ht]
\centering
\includegraphics[width=1\textwidth]{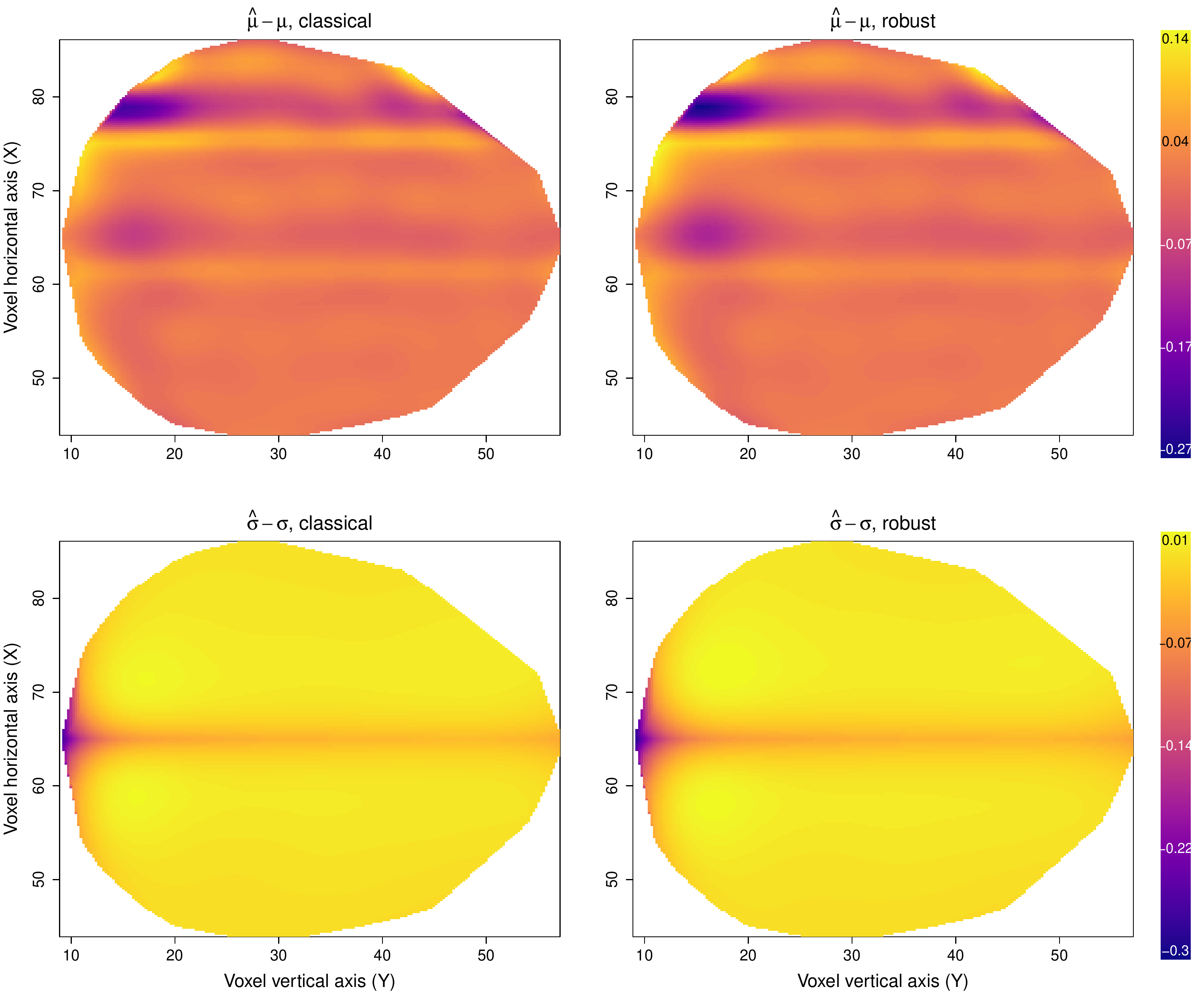}
\caption{GAMLSS simulation, surfaces of the average bias for the canonical parameters $\mu$ (top row) and $\sigma$ (bottom row) based on classical (left column) and robust (right column) estimation methods, at the assumed model.}
\label{fig:simgamlss.biasclean.musigma}
\end{figure}

\begin{figure}[!ht]
\centering
\includegraphics[width=1\textwidth]{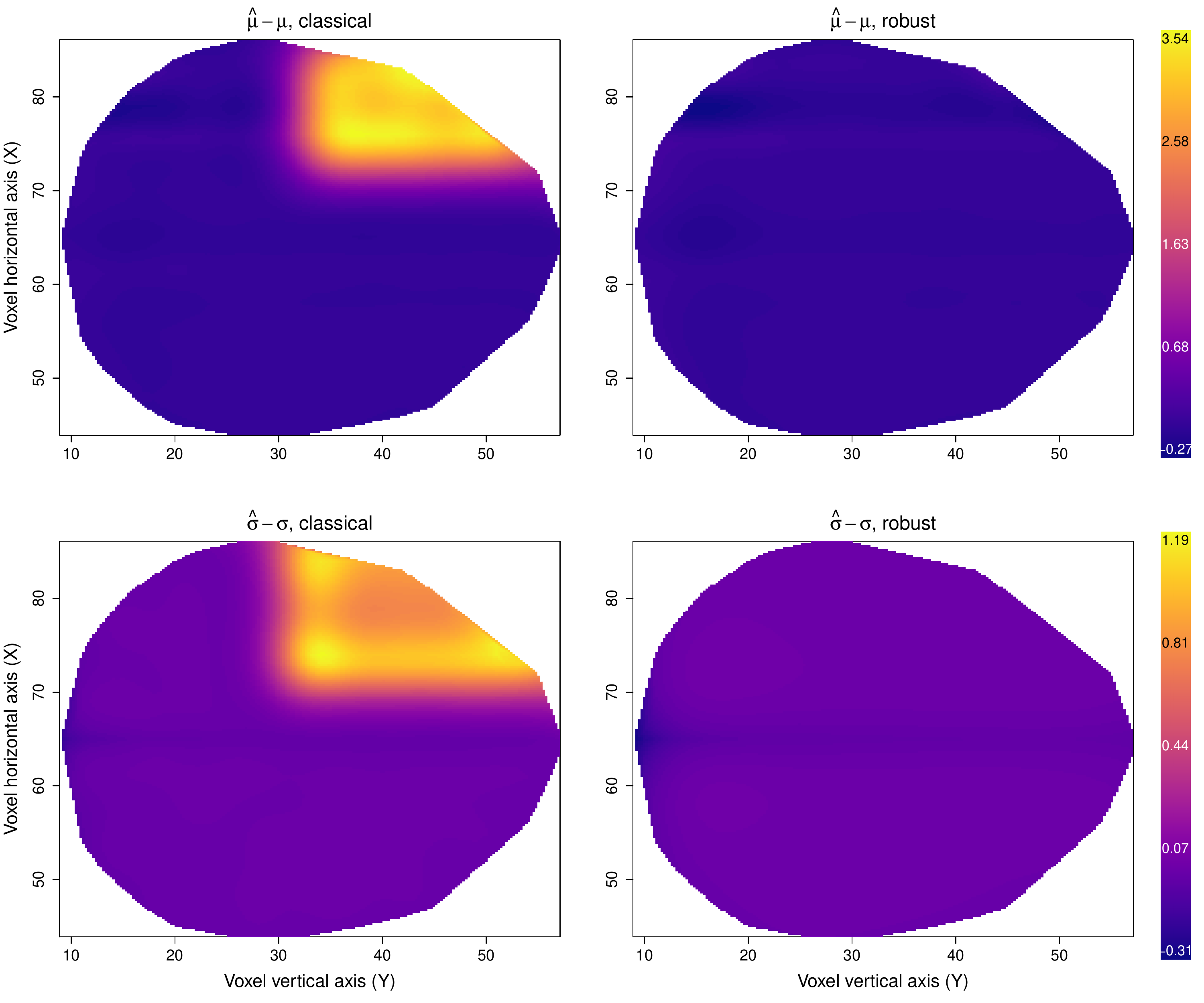}
\caption{GAMLSS simulation, surfaces of the average bias for the canonical parameters $\mu$ (top row) and $\sigma$ (bottom row) based on classical (left column) and robust (right column) estimation methods, under contmaination.}
\label{fig:simgamlss.biascont.musigma}
\end{figure}

\clearpage
\section*{Web Appendix~D: Supplementary Tables}

\renewcommand{\thetable}{S\arabic{table}} 
\setcounter{table}{0}                    


Table~S1 below lists the distributions implemented in the \texttt{R} package \texttt{GJRM}. These have been parametrized according to \citet{RigbyStasinopoulos2005} and are defined in terms of $\mu$, $\sigma$ and $\nu$. The means and variances of \texttt{DAGUM}, \texttt{FISK} (also known as log-logistic) and \texttt{SM} are indeterminate for certain values of $\sigma$ and $\nu$. If a parameter can only take positive values then the transformation/link function $\log(\cdot-\epsilon)$ is employed. If a parameter can take values in $(0,1)$ then the inverse of the cumulative distribution function of a standardized logistic is used. $B(\cdot,\cdot)$ is the beta function, $\Gamma(\cdot)$ is the gamma function, $\alpha = \sqrt{\frac{1}{\sigma^2} + \frac{2\mu}{\sigma}}$ and $K_\lambda(t) = \frac{1}{2} \int_0^\infty x^{\lambda-1}\exp\left\{ -0.5 t (x + x^{-1}) \right\}\,\td x$ is the modified Bessel function of the third kind. Argument \texttt{margin} of \texttt{gamlss()} in \texttt{GJRM} allows the user to employ the desired marginal distribution and can be set to any of the values within brackets next to the names of the distributions. In many cases the parameters of the distributions determine $\E[Y]$ and $\Va[Y]$ through functions thereof.

\begin{sidewaystable}
\scriptsize 
\centering
\begin{tabular}{lccccccccc}
\hline
 & $f(y|\mu, \sigma, \nu)$ & $\E(Y)$ & $\Va(Y)$ & \pbox{3cm}{\relax\ifvmode\centering\fi Support of $y$ and\\ Parameter ranges}   \\ 
\hline
beta (\texttt{"BE"})   & $\frac{y^{\alpha_1-1}(1-y)^{\alpha_2-1}}{B\left(\alpha_1,\alpha_2\right)}$ & $\mu$  & $\sigma^2\mu(1-\mu)$ & \pbox{4cm}{$ 0 < y < 1$\\$0 < \mu < 1, 0 < \sigma < 1$}   \\ 
Dagum (\texttt{"DAGUM"})         & $\frac{\sigma\nu}{y}\left[  \frac{\left(\frac{y}{\mu}\right)^{\sigma\nu}}{ \left\{\left(\frac{y}{\mu}\right)^{\sigma}+1\right\}^{\nu+1} }\right]$ & \pbox{4cm}{$-\frac{\mu}{\sigma}\frac{\Gamma\left(-\frac{1}{\sigma}\right)\Gamma\left(\frac{1}{\sigma}+\nu\right)}{\Gamma(\nu)}$\\$\text{if} \ \sigma > 1 $} & \pbox{4cm}{$-\left(\frac{\mu}{\sigma}\right)^2\left[2\sigma\frac{\Gamma\left(-\frac{2}{\sigma}\right)\Gamma\left(\frac{2}{\sigma}+\nu\right)}{\Gamma(\nu)}\right.$\\$\left.+\left\{\frac{\Gamma\left(-\frac{1}{\sigma}\right)\Gamma\left(\frac{1}{\sigma}+\nu\right)}{\Gamma(\nu)}\right\}^2  \right]$\\$\text{if} \ \sigma > 2$} & \pbox{4cm}{$y > 0$\\$\mu > 0, \sigma > 0, \nu > 0$}   \\
Fisk (\texttt{"FISK"})           & $\frac{\sigma y^{\sigma-1}}{\mu^\sigma\left\{ 1+\left(\frac{y}{\mu}\right)^\sigma\right\}^2}   $ & $ \frac{\mu\pi/\sigma}{\sin\left(\pi/\sigma\right)} \ \text{if} \ \sigma > 1   $ & \pbox{4cm}{$\mu^2\left\{ \frac{2\pi/\sigma}{\sin\left(2\pi/\sigma\right)}-\frac{\left(\pi/\sigma\right)^2}{\sin\left( \pi/\sigma \right)^2}  \right\}$\\$\text{if} \ \sigma > 2$} & \pbox{4cm}{$y > 0$\\$\mu > 0, \sigma > 0$}   \\
gamma (\texttt{"GA"})             & $\frac{1}{(\mu\sigma^2)^{\frac{1}{\sigma^2}}} \frac{y^{\frac{1}{\sigma^2}-1}\exp\left(-\frac{y}{\mu\sigma^2}\right)}{\Gamma\left(\frac{1}{\sigma^2}\right)}$ & $\mu$ & $\mu^2\sigma^2$ & \pbox{4cm}{$y > 0$\\$\mu > 0, \sigma > 0$}   \\
Gumbel (\texttt{"GU"})            & $\frac{1}{\sigma} \exp\left\{\left(\frac{y-\mu}{\sigma}\right)-\exp\left(\frac{y-\mu}{\sigma}\right) \right\}  $ & $\mu - 0.57722\sigma$  & $\frac{\pi^2\sigma^2}{6}$ & \pbox{4cm}{$ -\infty < y < \infty$\\$-\infty < \mu < \infty, \sigma > 0$}   \\ 
inverse Gaussian (\texttt{"iG"})  & $\frac{1}{\sqrt{2\pi\sigma^2y^3}}\exp\left\{ -\frac{1}{2\mu^2\sigma^2y}\left(y-\mu\right)^2 \right\}$ & $\mu$ & $\mu^3\sigma^2$ & \pbox{4cm}{$y > 0$\\$\mu > 0, \sigma > 0$}   \\
log-normal (\texttt{"LN"})        & $\frac{1}{y\sigma\sqrt{2\pi}}\exp\left[-\frac{\left\{\log(y)-\mu\right\}^2}{2\sigma^2}\right]$ & $\sqrt{\exp\left(\sigma^2\right)}\exp\left(\mu\right)$ & \pbox{4cm}{$\exp\left(\sigma^2\right)\left\{\exp\left(\sigma^2\right)\right.$\\$\left.-1\right\}\exp\left(2\mu\right)$} & \pbox{4cm}{$y > 0$\\$-\infty < \mu < \infty, \sigma > 0$} \\
logistic (\texttt{"LO"})          & $\frac{1}{\sigma}\left\{\exp\left(-\frac{y-\mu}{\sigma}\right)\right\}\left\{1+\exp\left(-\frac{y-\mu}{\sigma}\right)\right\}^{-2}$ & $\mu$ & $\frac{\pi^2\sigma^2}{3}$  & \pbox{4cm}{$ -\infty < y < \infty$\\$-\infty < \mu < \infty, \sigma > 0$}\\
normal (\texttt{"N"})            & $\frac{1}{\sigma\sqrt{2\pi}}\exp\left\{-\frac{\left(y-\mu\right)^2}{2\sigma^2}\right\}$ & $\mu$ & $\sigma^2$ & \pbox{4cm}{$ -\infty < y < \infty$\\$-\infty < \mu < \infty, \sigma > 0$}\\
reverse Gumbel (\texttt{"rGU"})   & $\frac{1}{\sigma} \exp\left\{\left(-\frac{y-\mu}{\sigma}\right)-\exp\left(-\frac{y-\mu}{\sigma}\right) \right\}  $ & $\mu + 0.57722\sigma$  & $\frac{\pi^2\sigma^2}{6}$ & \pbox{4cm}{$ -\infty < y < \infty$\\$-\infty < \mu < \infty, \sigma > 0$}  \\
Singh-Maddala (\texttt{"SM"})   & $\frac{\sigma\nu y^{\sigma-1}}{\mu^\sigma\left\{1+\left(\frac{y}{\mu}\right)^\sigma\right\}^{\nu+1}}$ & \pbox{4cm}{$\mu\frac{\Gamma\left(1+\frac{1}{\sigma}\right)\Gamma\left(-\frac{1}{\sigma}+\nu\right)}{\Gamma(\nu)}$\\$\text{if} \ \sigma\nu > 1 $} & \pbox{4.5cm}{$\mu^2\left\{\Gamma\left(1+\frac{2}{\sigma}\right)\Gamma(\nu)\Gamma\left(-\frac{2}{\sigma}+\nu\right)\right.$\\$\left.-\Gamma\left(1+\frac{1}{\sigma}\right)^2\Gamma\left(-\frac{1}{\sigma}+\nu\right)^2 \right\}$\\$\text{if} \ \sigma\nu > 2$} & \pbox{4cm}{$y > 0$\\$\mu > 0, \sigma > 0, \nu > 0$}   \\
Weibull (\texttt{"WEI"})         & $\frac{\sigma}{\mu}\left(\frac{y}{\mu}\right)^{\sigma-1}\exp\left\{-\left(\frac{y}{\mu}\right)^\sigma\right\}$ & $\mu\Gamma\left(\frac{1}{\sigma}+1\right)$ & \pbox{4cm}{$\mu^2\left[\Gamma\left(\frac{2}{\sigma}+1\right)\right.$\\$\left.-\left\{\Gamma\left(\frac{1}{\sigma}+1\right)\right\}^2\right]$} & \pbox{4cm}{$y > 0$\\$\mu > 0, \sigma > 0$}  \\
Poisson (\texttt{"PO"})    & $\frac{\exp(-\mu)\mu^y}{y!}$ & $\mu$ & $\mu$ & \pbox{4cm}{$y = 0,1,\ldots$\\$\mu > 0$}  \\
negative binomial type I (\texttt{"NBI"})    & $\frac{\Gamma(y+1/\sigma)}{\Gamma(1/\sigma)\Gamma(y+1)} \left(\frac{\sigma\mu}{1+\sigma\mu}\right)^y \left(\frac{1}{1+\sigma\mu}\right)^{1/\sigma}$   & $\mu$ & $\mu + \sigma \mu^2$ & \pbox{4cm}{$y = 0,1,\ldots$\\$\mu > 0, \sigma > 0$}  \\
Poisson-inverse Gaussian (\texttt{"PIG"})    & $ \left(\frac{2\alpha}{\pi}\right)^{0.5} \frac{\mu^y\exp(1/\sigma)K_{y-0.5}(\alpha)}{(\alpha \sigma)^y y!} $ & $\mu$ & $\mu + \sigma \mu^2$ & \pbox{4cm}{$y = 0,1,\ldots$\\$\mu > 0, \sigma > 0$}  \\
negative binomial type II (\texttt{"NBII"})    & $\frac{\Gamma(y + \mu/\sigma)\sigma^y}{\Gamma(\mu/\sigma)\Gamma(y+1)(1+\sigma)^{y+\mu/\sigma}}$  & $\mu$ & $(1 + \sigma)\mu$ & \pbox{4cm}{$y = 0,1,\ldots$\\$\mu > 0, \sigma > 0$}  \\
zero-truncated Poisson (\texttt{"ZTP"})    & $\frac{\mu^y}{y!(1-\exp(-\mu))}$ & $\frac{\mu}{1-\exp(-\mu)}$ & $\frac{\mu(1-\exp(-\mu)(\mu+1))}{(1-\exp(-\mu))^2}$ & \pbox{4cm}{$y = 1,2, \ldots$\\$\mu > 0$}  \\
\hline \\ \\
\end{tabular}
\caption{Definition and some properties of the distributions implemented in \texttt{GJRM}.}
\end{sidewaystable}

\clearpage 
\bibliographystyle{plainnat} 
\bibliography{RobustGAMLSSbiblio}



\end{document}